\documentclass[11pt]{article}

\usepackage[letterpaper, margin=1in]{geometry}
\usepackage{amssymb}
\usepackage{amsmath}
\usepackage{amsthm}
\usepackage{newclude}
\usepackage{lmodern}
\usepackage[T1]{fontenc}
\usepackage{microtype}
\usepackage{algorithm}
\usepackage{algorithmicx}
\usepackage{algpseudocode}
\usepackage{fixltx2e}
\usepackage{color}
\usepackage{graphicx}
\usepackage[hidelinks]{hyperref}
\usepackage{subcaption}
\usepackage{multicol}
\usepackage{setspace}

\DeclareMathOperator{\poly}{poly}

\theoremstyle{plain}
\newtheorem{theorem}{Theorem}[section]
\newtheorem{lemma}[theorem]{Lemma}

\newtheorem{proposition}[theorem]{Proposition}
\newtheorem{claim}[theorem]{Claim}
\newtheorem{remark}[theorem]{Remark}

\theoremstyle{definition}
\newtheorem{definition}[theorem]{Definition}

\MakeRobust{\Call}
\algtext*{EndWhile}
\algtext*{EndFor}
\algtext*{EndIf}
\algtext*{EndFunction}

\newcommand{\Rc}{\mathcal{R}}
\newcommand{\Nbb}{\mathbb{N}}
\newcommand{\Rbb}{\mathbb{R}}

\newcommand{\eps}{\varepsilon}
\newcommand{\set}[1]{\left\{#1\right\}}
\newcommand{\Prb}[2]{\mathrm{Pr}_{#1}\left[#2\right]}
\newcommand{\Exp}[2]{\mathrm{E}_{#1}\left[#2\right]}
\DeclareMathOperator{\supp}{supp}

\newcommand{\hl}[1]{}
\newcommand{\aanote}[1]{}

\begin{document}
\begin{titlepage}
    \title{Optimal Data-Dependent Hashing for Approximate Near Neighbors}
    \author{Alexandr Andoni\footnote{Work was done in part while both authors were at Microsoft Research Silicon Valley.}{\enskip}\footnote{Simons Institute for the Theory
        of Computing, UC Berkeley, \texttt{andoni@mit.edu}} \and Ilya Razenshteyn%
\footnotemark[1]{\enskip}\footnote{CSAIL MIT, \texttt{ilyaraz@mit.edu}}}
    \maketitle
    \thispagestyle{empty}
\begin{abstract}

We show an optimal data-dependent hashing scheme for the
approximate near neighbor problem. For an $n$-point dataset in a $d$-dimensional space
our data structure achieves query time $O(d \cdot
n^{\rho+o(1)})$ and space $O(n^{1+\rho+o(1)} + d \cdot n)$, where
$\rho=\tfrac{1}{2c^2-1}$ for the Euclidean space and approximation
$c>1$. For the Hamming space, we obtain an exponent of
$\rho=\tfrac{1}{2c-1}$.

Our result completes the direction set forth in \cite{ainr-blsh-14}
who gave a proof-of-concept that data-dependent hashing can outperform
classical Locality Sensitive Hashing (LSH). In contrast to \cite{ainr-blsh-14},
the new bound is not only optimal, but in fact improves over the
best (optimal) LSH data structures \cite{im-anntr-98,ai-nohaa-06} for \emph{all}
approximation factors $c>1$.

From the technical perspective, we proceed by decomposing
an \emph{arbitrary} dataset into several subsets that are, in a
certain sense, \emph{pseudo-random}.
\end{abstract}

\end{titlepage}
\tableofcontents

\onehalfspacing
  
    \section{Introduction}

In the near neighbor search problem, we are given a set $P$ of $n$
points in a $d$-dimensional space, and the goal is to
build a data structure that, given
a query point $q$, reports any point within a given distance $r$ to
the query.  The problem is of major importance in several areas, such
as databases, data mining, information retrieval, computer vision,
computational geometry, signal processing, etc.

Efficient near(est) neighbor algorithms are known for the case when
the dimension $d$ is ``low'' (e.g., see~\cite{c-racpq-88, m-plah-93}).
However, the current solutions suffer from ``the curse of
dimensionality'' phenomenon: either space or query time are
exponential in the dimension $d$. To escape this curse, researchers
proposed {\em approximation} algorithms for the problem.  In the $(c,
r)$-approximate near neighbor problem (ANN), the data structure may
return any data point whose distance from the query is at most $c r$,
for an approximation factor $c > 1$ (provided that there exists a data
point within distance $r$ from the query).  Many approximation algorithms for the
problem are known:
e.g., see surveys~\cite{AI-CACM, a-nnson-09}.

To address the ANN problem, Indyk and Motwani proposed the {\em
  Locality Sensitive Hashing} scheme (LSH), which has since proved to
be influential in theory and practice
\cite{im-anntr-98,him-anntr-12}. In particular, LSH yields the best
ANN data structures for the regime of sub-quadratic space and constant
approximation factor, which turns out to be the most important regime
from the practical perspective.  The main idea is to hash the points
such that the probability of collision is much higher for points which
are close to each other (at distance $\le r$) than for those which are
far apart (at distance $\ge cr$). Given such hash functions, one can
retrieve near neighbors by hashing the query point and retrieving
elements stored in buckets containing that point.  If the probability
of collision is at least $p_1$ for the close points and at most $p_2$
for the far points, the algorithm solves the $(c, r)$-ANN using
$n^{1+\rho}$ extra space and $dn^{\rho}$ query time\footnote{For
  exposition purposes, we are suppressing the time to compute hash
  functions, which we assume to require $n^{o(1)}$ time and $n^{o(1)}$
  space. We also assume distances can be computed in $O(d)$ time, and
  that $1/p_1 =n^{o(1)}$.}, where
$\rho=\log(1/p_1)/\log(1/p_2)$~\cite{him-anntr-12}. The value of the
exponent $\rho$ thus determines the ``quality'' of the LSH families
used.

A natural question emerged: what is the best possible exponent
$\rho$?  The original LSH paper \cite{im-anntr-98} showed $\rho\le 1/c$
for both Hamming and Euclidean spaces. Focusing on the Euclidean
space, subsequent research showed that one
can obtain a better exponent: $\rho\le 1/c^2$ \cite{diim-lshsp-04,
  ai-nohaa-06}\footnote{Ignoring terms vanishing with $n$; the exact
  dependence is $\rho=1/c^2+1/\log^{\Omega(1)}n$.}. Complementing
these results, lower bounds on $\rho$ showed that this bound is
tight \cite{mnp-lblsh-07, owz-olbls-11}, thus settling the question: the
best exponent is $\rho=1/c^2$ for the Euclidean space.

\paragraph{Data-dependent hashing.} 
Surprisingly, while the best possible LSH exponent $\rho$ has been
settled, it turns out there exist {\em more efficient} ANN data
structures, which step outside the LSH framework. In particular,
\cite{ainr-blsh-14} obtain the exponent of
$\rho=\tfrac{7/8}{c^2}+\tfrac{O(1)}{c^{3}}$ by considering {\em
  data-dependent hashing}, i.e., a randomized hash family that itself
depends on the actual points in the dataset. We stress that this
approach gives improvement for {\em worst-case} datasets, which is
somewhat unexpected. To put this into a perspective: if one were to
assume that the dataset has some {\em special structure}, it would be
more natural to expect speed-ups with data-dependent hashing: such
hashing may adapt to the special structure, perhaps implicitly, as was
done in, say, \cite{df-rptldm-08, vkd-wspta-09, aakk-spectral-14}.
However, in our setting there is no assumed structure to adapt to, and
hence it is unclear why data-dependent hashing shall help. (To compare
with classical, non-geometric hashing, the most similar situation where
data-dependent hashing helps in the worst-case seems to be the perfect
hashing~\cite{fks-sstwc-84}.) Note that for the case of Hamming space,
\cite{ainr-blsh-14} has been the first and only improvement over
\cite{im-anntr-98} since the introduction of LSH (see Section
\ref{sec:implications}).

Thus the core question resurfaced: what is the best possible exponent
for {\em data-dependent} hashing schemes? To formulate the question
correctly, we also need to require that the hash family is ``nice'':
otherwise, the trivially best solution is the Voronoi diagram of $n$
points at hand, which is obviously useless (computing the hash
function is as hard as the original problem!). A natural way to
preclude such non-viable solutions is to require that each hash function
can be ``efficiently described'', i.e., it has a description of
$n^{1 - \Omega(1)}$ bits (e.g., this property is satisfied by all the LSH
functions we are aware~of).

\subsection{Main result}
We present an optimal data-dependent hash
family that achieves the following exponent\footnote{Again, we ignore the
  additive term that vanishes with $n$.} for ANN under the Euclidean distance:
\begin{equation}
\label{eqn:rhoBound}
\rho=\frac{1}{2c^2-1}.
\end{equation}

Specifically, we obtain the following main theorem.
\begin{theorem}
For fixed approximation $c>1$ and threshold $r>0$, one can solve the
$(c,r)$-ANN problem in $d$-dimensional Euclidean space on $n$ points with
$O(d \cdot n^{\rho+o(1)})$ query time, $O(n^{1+\rho+o(1)}+d\cdot n)$ space, and
$O(d \cdot n^{1+\rho+o(1)})$ preprocessing
time, where $\rho=\tfrac{1}{2c^2-1}$.
\end{theorem}

\hl{Needs some work (up until the end of the section)}

The optimality of our bound~\eqref{eqn:rhoBound} is established in a separate paper~\cite{ar-ddLSHlb-15}.
There we build on a \emph{data-independent} lower bound for a \emph{random dataset}~\cite{mnp-lblsh-07}.
First, we improve upon~\cite{mnp-lblsh-07} quantitatively and obtain the lower bound $\rho \geq \frac{1}{2c^2 - 1} - o(1)$
for the data-independent case, thus providing an alternative self-contained proof of the lower bound from~\cite{d-bcits-10}.
Second, we argue that if there is a good data-dependent hashing family for a random dataset
with hash functions having low description complexity,
then it can be converted into a data-independent family.
Hence, the lower bound $\rho \geq \frac{1}{2c^2 - 1} - o(1)$ applies to the data-dependent case as well.
For the details, we refer readers to~\cite{ar-ddLSHlb-15}.

An important aspect of our algorithm is that it effectively reduces
ANN on a generic dataset to ANN on an (essentially) random
dataset.
The latter is the most natural ``hard distribution'' for
the ANN problem. Besides the aforementioned lower bounds, it is also a source of
\emph{cell-probe} lower bounds for
ANN~\cite{ptw-galba-08,ptw-lbnns-10}. Hence, looking forward, to
further improve the efficiency of ANN, one would have first to improve
the random dataset case, which seems to require fundamentally
different techniques, if possible at all.

The importance of this reduction can be seen from the progress on the
closest pair problem by Valiant \cite{v-fcsta-12}. In particular,
Valiant gives an algorithm with $n^{1.62}\cdot \poly(\tfrac{d}{c-1})$
runtime for the aforementioned {\em random instances}.\footnote{Note
  that this improves over \cite{d-bcits-10}: Valiant exploits fast
  matrix multiplication algorithms to step outside Dubiner's hashing
  framework altogether.}  Obtaining similar runtime for the {\em worst
  case} (e.g., via a similar reduction) would {\em refute the Strong
  Exponential-Time Hypothesis} (SETH).

We also point out that---besides achieving the optimal bound---the
new algorithm has two further advantages over the one
from~\cite{ainr-blsh-14}.  First, our bound~\eqref{eqn:rhoBound} is
better than the optimal LSH bound $1 / c^2$ \emph{for every} $c > 1$
(the bound from~\cite{ainr-blsh-14} is only better for sufficiently
large~$c$).  Second, the preprocessing time of our algorithm is near-linear
in the amount of space used, improving over the quadratic 
preprocessing time of~\cite{ainr-blsh-14}.

\subsection{Techniques}
\label{sec:techniques}

The general approach is via data-dependent LSH families, which can be
equivalently seen as data-dependent random {\em space
  partitions}. Such space partitions are usually constructed
iteratively: first we partition the space very coarsely, then we
refine the partition iteratively a number of times. In standard LSH,
each iteration of partitioning is random i.i.d., and the overall data
structure consists of $n^\rho$ such iterative space partitions,
constructed independently (see \cite{him-anntr-12} for details).

\hl{The following paragraph strongly feels out of place!}

For the latter discussion, it is useful to keep in mind what are the
\emph{random dataset instances} for ANN.  Consider a sphere of radius
$cr / \sqrt{2}$ in $\Rbb^{d}$ for $d=1000 \log n$. The data set is
obtained by sampling $n$ points on the sphere uniformly at random. To
generate a query, one chooses a data point uniformly at random, and
plants a query at distance at most within $r$ from it uniformly at
random.  It is not hard to see that with very high probability, the
query will be at least $cr - o(1)$ apart from all the data points
except one. Thus, any data structure for $(c - o(1), r)$-ANN must be
able to recover the data point the query was planted to.

Let us first contrast our approach to the previous algorithm of
\cite{ainr-blsh-14}. That result improved over LSH by identifying a
``nice configuration'' of a dataset, for which one can design a hash
family with better $\rho<1/c^2$. It turns out that the right notion
of niceness is the ability to enclose dataset into a ball of small
radius, of order $O(cr)$ (the aforementioned random instance
corresponds to ``tightest possible'' radius of $cr / \sqrt{2}$).
Moreover, the smaller the enclosing ball is, the better the exponent
$\rho$ one can obtain.  The iterative partitioning from
\cite{ainr-blsh-14} consists of two rounds. During the
first round, one partitions the space so that the dataset in each part
would be ``low-diameter'' with high probability.  This step uses
classical, data-independent LSH, and hence effectively has quality
$\rho=1/c^2$. During the second round, one would apply
``low-diameter'' LSH with quality $\rho<1/c^2$. The final exponent
$\rho$ is a weighted average of qualities of the two rounds. While one
can generalize their approach to any number of rounds, the best
exponent $\rho$ one can obtain this way is around $0.73/c^2+O(1/c^3)$ \cite{r-msthesis-14},
which falls short of~(\ref{eqn:rhoBound}).

In fact, \cite{ainr-blsh-14} cannot obtain the optimal $\rho$ as
in~\eqref{eqn:rhoBound} {\em in principle}.
The fundamental issue is that, before one completes the reduction to a
``nice configuration'', one must incur some ``waste''. In particular,
the first round uses (non-optimal) data-independent hashing, and hence
the ``average'' $\rho$ cannot meet the best-achievable
$\rho$. (Moreover, even the second round of the algorithm does not
achieve the optimal $\rho$.)

Thus, the real challenge remained: how to perform {\em each} iteration
of the partitioning with {\em optimal~$\rho$}? E.g., we must succeed even
during the very first iteration, on a dataset without any
structure whatsoever.

Our new algorithm resolves precisely this challenge.  For simplicity,
let us assume that all the data points lie on a sphere of radius
$R$. (It is helpful to think of $R$ as being, say, $1000 cr$: e.g.,
the low-diameter family from~\cite{ainr-blsh-14} gives almost no
advantage over the data-independent $\rho = 1/c^2$ for such $R$).  We
start by decomposing the data set into a small number of \emph{dense
  clusters} (by this we mean a spherical cap that is slightly smaller
than a hemisphere and that covers $n^{1 - o(1)}$ points) and a
\emph{pseudo-random} remainder that has no dense parts. For dense
clusters we recurse enclosing them in balls of radius slightly smaller
than $R$, and for the pseudo-random part we just apply one iteration of
the ``low-diameter'' LSH from~\cite{ainr-blsh-14} and then recurse on
each part.  This partitioning subroutine makes progress in two ways.
For dense clusters we slightly reduce the radius of the
instance. Thus, after a bounded number of such reductions we will
arrive to an instance that can be easily handled using the
low-diameter family.  For the pseudo-random remainder, we can argue
that the low-diameter family works ``unreasonably'' well: intuitively, it
follows from the fact that almost all pairs of data points are very
far apart (roughly speaking, at distance almost $\sqrt{2}
R$). We call the remainder pseudo-random precisely because of the latter
property: random points on a sphere of radius $R$
are essentially $(\sqrt{2} R)$-separated.

We note that one can see the partition from
above as a (kind of) decision tree, albeit with a few particularities.
Each node of the decision tree has a number of children that partition
the dataset points (reaching this node), corresponding to the dense
clusters as well as to all the (non-empty) parts of the aforementioned
``low-diameter'' LSH partition. In fact, it will be necessary for some
points to be replicated in a few children (hence the decision tree
size may be $\gg n$), though we will show that the degree of
replication is bounded (a factor of $n^{o(1)}$ overall). The query
procedure will search the decision tree by following a path from the
root to the leaves. Again, it may be necessary to follow a few
children at a decision node at a time; but the replication is bounded
as well. The final data structure is composed of $n^\rho$ such
decision trees, and the query algorithm queries each of them.

A new aspect of our analysis is that, among other things, we will need to
understand how the low-diameter family introduced in~\cite{ainr-blsh-14} works
on \emph{triples of points}: without this analysis, one can only get a
bound of
$$
    \rho = \frac{1}{2c^2 - 2},
$$ which is much worse than~(\ref{eqn:rhoBound}) for small $c$.  To
    the best of our knowledge, this is the first time when one needs
    to go beyond the ``pairwise'' analysis in the study of LSH.

\subsection{Further implications and connections}

\label{sec:implications}

Our algorithm also directly applies to the Hamming metric,
for which it achieves\footnote{This follows from a standard embedding of
the $\ell_1$ norm into
$\ell_2$-squared~\cite{llr-ggsaa-95}.} an exponent of
$$\rho=\frac{1}{2c-1}.
$$ This is a nearly quadratic improvement over the original LSH paper
\cite{im-anntr-98}, which obtained exponent $\rho=1/c$, previously
shown to be optimal for the classical LSH in \cite{owz-olbls-11}. The
result of \cite{ainr-blsh-14} was the first to bypass the 1998 bound
via data-dependent hashing, achieving
$\rho=\tfrac{7/8}{c}+\tfrac{O(1)}{c^{3/2}}$, an improvement for large
enough $c$. Our new bound improves over \cite{im-anntr-98} for all
values of $c$, and is also optimal (the above discussion for $\ell_2$
applies here as well).

From a broader perspective, we would like to point out the related
developments in practice.
Many or most of the practical applications of LSH involve designing
{\em data-aware} hash
functions~\cite{s-rnnsk-91,m-fnnab-01,vkd-wspta-09,wtf-sh-08,sh-sh-09,ysrl-pcr-11}
(see also a recent survey~\cite{wssj-hsss-14}).  The challenge of
understanding and exploiting the relative strengths of data-oblivious
versus data-aware methods has been recognized as a major open question
in the area (e.g., see~\cite{fmda-13}, page 77).  This paper can be
seen as part of the efforts addressing the challenge.

Let us also point a recent paper~\cite{practicalballcarving} that
provides a practical analog of Spherical LSH (see Section~\ref{sec_glsh}) that
not only has the same theoretical guarantees, but is also practical; in particular, it outperforms in practice
celebrated Hyperplane LSH~\cite{c-setra-02} for similarity search on a sphere.

Finally, we note that our \emph{inherently static} data structure can be dynamized: we can perform insertions/deletions in time
$d \cdot n^{\frac{1}{2c^2 - 1} + o(1)}$ using a well-known \emph{dynamization} technique for decomposable search problems~\cite{ovl81}.
Here we crucially use the fast preprocessing routine developed in the present paper.

    \section{Preliminaries}
In the text we denote the $\ell_2$ norm by $\| \cdot \|$. From now on, when we use $O(\cdot)$, $o(\cdot)$,
$\Omega(\cdot)$ or $\omega(\cdot)$ we explicitly write all the parameters that the corresponding constant factors
depend on as subscripts (the variable is always $n$ or derived
functions of $n$).
Our main tool will be random partitions of a metric space. 
For a partition $\Rc$ and a point $p$ we denote $\Rc(p)$ the part of $\Rc$, which $p$ belongs to.
If $\Rc$ is a partition of a \emph{subset} of the space, then we denote $\bigcup \Rc$ the union of all
the pieces of $\Rc$.
By $N(a, \sigma^2)$ we denote a standard Gaussian with mean $a$ and variance $\sigma^2$.
We denote the closed Euclidean ball with a center $u$ and a radius $r \geq 0$ by $B(u, r)$.
By $\partial B(u, r)$ we denote the corresponding \emph{sphere}.
We denote $S^{d-1} \subset \Rbb^d$ the unit Euclidean sphere in $\Rbb^d$ with the center being the origin.
A spherical cap can be considered as a ball on a sphere with metric inherited from the ambient Euclidean
distance. We define a \emph{radius} of a spherical cap to be the radius of the corresponding ball.
For instance, the radius of a hemisphere of a unit sphere is equal to $\sqrt{2}$.

\begin{definition}
    The \emph{$(c, r)$-Approximate Near Neighbor problem (ANN)} with failure
    probability $f$ is to construct a data structure over a set
    of points $P \subseteq \Rbb^d$ supporting the following
    query: given any fixed query point $q \in \Rbb^d$, if there exists
    $p \in P$ with $\|p - q\| \leq r$, then report some $p' \in P$
    such that $\|p' - q\| \leq cr$, with probability at least $1 - f$.
\end{definition}
Note that we allow preprocessing to be randomized as well,
and we measure the probability of success over the random coins tossed
during \emph{both} preprocessing and query phases.

\begin{definition}[\cite{him-anntr-12}]
    We call a random partition $\Rc$ of $\Rbb^d$
    \emph{$(r_1, r_2, p_1, p_2)$-sensitive}, if
    for every $x, y \in X$ we have
    $\Prb{\Rc}{\Rc(x) = \Rc(y)} \geq p_1$ if $\|x - y\| \leq r_1$,
    and $\Prb{\Rc}{\Rc(x) = \Rc(y)} \leq p_2$ if $\|x - y\| \geq r_2$.
\end{definition}

\paragraph{Remark:} For $\mathcal{\Rc}$ to be useful we require that $r_1 < r_2$ and $p_1 > p_2$.

Now we are ready to state a very general way to solve ANN, if we have
a good $(r, cr, p_1, p_2)$-sensitive partition~\cite{im-anntr-98,
  him-anntr-12}.  The following theorem gives a data structure with
near-linear space and small query time, but with probability
of success being only inversely polynomial in the number of points.

\begin{theorem}[\cite{im-anntr-98, him-anntr-12}]
    \label{lsh_to_nn}
    Suppose that there is a $(r, cr, p_1, p_2)$-sensitive partition $\Rc$
    of $\Rbb^d$, where $(p_1, p_2) \in (0, 1)$
    and let $\rho = \ln(1 / p_1) / \ln(1 / p_2)$.
    Assume that $p_1, p_2 \geq 1 / n^{o_c(1)}$, one can sample a partition from $\Rc$ in time $n^{o_c(1)}$,
    store it in space $n^{o_c(1)}$ and perform point location in time $n^{o_c(1)}$.
    Then there exists a data structure for $(c, r)$-ANN over a set
    $P \subseteq \Rbb^d$ with $|P| = n$ with preprocessing time $O(dn^{1+o_c(1)})$,
    probability of success at least $n^{-\rho - o_c(1)}$, space consumption (in addition to the data points)
    $O(n^{1 + o_c(1)})$ and \emph{expected} query time $O(dn^{o_c(1)})$.
\end{theorem}

\begin{remark} 
\label{rem:fullDS}
To obtain the final data structure we
increase the probability of success from $n^{-\rho - o_c(1)}$ to
$0.99$ by building $O(n^{\rho + o_c(1)})$ independent data
structures from Theorem~\ref{lsh_to_nn}. Overall, we obtain
$O(dn^{\rho + o_c(1)})$ query time, $O(dn + n^{1 + \rho + o_c(1)})$
space, and $O(dn^{1 + \rho + o_c(1)})$ preprocessing time.
\end{remark}

In our algorithm, we will also be using the following (data-independent) LSH scheme.
\begin{theorem}[\cite{diim-lshsp-04}]
    \label{diim}
    There exists a random partition $\Rc$ of $\Rbb^d$
    such that for every $u, v \in \Rbb^d$ with $\|u - v\| = \tau$
    one has
    $
        \ln \tfrac{1}{\Prb{\Rc}{\Rc(u) = \Rc(v)}}
        =
        \left(1 + O_{\tau}\left(1/d\right)\right) \cdot \tau \sqrt{d}.
    $
    Moreover, $\Rc$ can be sampled in time $d^{O(1)}$, stored in space
    $d^{O(1)}$ and a point can be located in $\Rc$ in time $d^{O(1)}$.
\end{theorem}

    \section{Spherical LSH}
\label{sec_glsh}

In this section, we describe a partitioning scheme of the unit sphere
$S^{d-1}$, termed \emph{Spherical LSH}. We will use Spherical LSH in
our data structure described in the next section. While the Spherical
LSH was introduced in~\cite{ainr-blsh-14}, we need to show a new
important property of it.
We then illustrate how Spherical LSH achieves optimal $\rho$ for the
ANN problem in the ``base case'' of {\em random} instances. As
mentioned in the Introduction, the main thrust of the new data
structure will be to reduce a worst-case dataset to this ``base case''.
Let us point out that a partitioning procedure similar to Spherical LSH has been used
in~\cite{kms-agcsp-98} for completely different purpose.

The main idea of the Spherical LSH is to ``carve''
spherical caps of radius $\sqrt{2} - o(1)$ (\emph{almost
  hemispheres}). The partitioning proceeds as follows:
\begin{algorithmic}
    \State $\Rc \gets \emptyset$
    \While{$\bigcup \Rc \ne S^{d-1}$}
        \State sample $g \sim N(0, 1)^d$
        \State $U \gets \set{u \in S^{d-1} \colon
                \langle u, g \rangle \geq d^{1/4}} \setminus
                \bigcup \Rc$
        \If{$U \ne \emptyset$}
            \State $\Rc \gets \Rc \cup \set{U}$
        \EndIf
    \EndWhile
\end{algorithmic}
Here $\Rc$ denotes the resulting partition of $S^{d-1}$, $\bigcup \Rc$
denotes the union of all elements of $\Rc$ (the currently partitioned
subset of $S^{d-1}$) and $N(0, 1)^d$ is a standard $d$-dimensional
Gaussian.

This partitioning scheme is not efficient: we can not quickly compute
$\bigcup \Rc$, and the partitioning process can potentially be infinite.
We will address these issues in Section~\ref{sec_eff}.

Now let us state the properties of Spherical LSH (the efficiency aspects are for a slightly modified variant of
it, according to Section~\ref{sec_eff}). The proof is deferred to Appendix~\ref{app_glsh}.

\begin{theorem}
    \label{gaussian_lsh_ideal}
    There exists a positive constant $\delta > 0$ such that
    for sufficiently large $d$, Spherical LSH $\Rc$ on $S^{d-1}$ has the
    following properties.
    Suppose that $\eps=\eps(d)>0$ tends to $0$ as $d$ tends to infinity;
    also let
    $\tau \in [d^{-\delta}; 2 - d^{-\delta}]$.
    Then, for every $u, v, w \in S^{d-1}$, we have
            \begin{equation}
                \label{glsh_lb}
                \|u - v\| \geq \tau \mbox{ implies }
                \ln \frac{1}{\Prb{\Rc}{\Rc(u) = \Rc(v)}} \geq
                (1 - d^{-\Omega(1)}) \cdot \frac{\tau^2}
                {4 - \tau^2}
                \cdot \frac{\sqrt{d}}{2},
            \end{equation}
            \begin{equation}
                \label{glsh_ub}
                \|u - v\| \leq \tau \mbox{ implies }
                \ln \frac{1}{\Prb{\Rc}{\Rc(u) = \Rc(v)}} \leq
                (1 + d^{-\Omega(1)}) \cdot \frac{\tau^2}
                {4 - \tau^2}
                \cdot \frac{\sqrt{d}}{2},
            \end{equation}
            \begin{multline}
                \label{glsh_cond}
                \|u - w\|, \|v - w\| \in \sqrt{2} \pm \eps \mbox{ and } \|u - v\| \leq 1.99 \\\mbox{ imply }
                \ln \frac{1}{\Prb{\Rc}{\Rc(u) = \Rc(w) \mid \Rc(u) = \Rc(v)}} \geq
                (1 - \eps^{\Omega(1)} - d^{-\Omega(1)})
                \cdot \frac{\sqrt{d}}{2}.
            \end{multline}
    Moreover, we can sample a partition in time $\exp(O(\sqrt{d}))$, store it in space $\exp(O(\sqrt{d}))$
    and locate a point from $S^{d-1}$ in it in time $\exp(O(\sqrt{d}))$.
\end{theorem}

\paragraph{Discussion of the three-point collision property \eqref{glsh_cond}.}
While properties \eqref{glsh_lb} and~\eqref{glsh_ub} were derived
in~\cite{ainr-blsh-14} (under an additional assumption $\tau \leq
\sqrt{2}$), the property \eqref{glsh_cond}
is the new contribution of Theorem~\ref{gaussian_lsh_ideal}.

Let us elaborate why proving~\eqref{glsh_cond} is challenging. First, one
can easily show that
\begin{align}
    \ln \frac{1}{\Prb{\Rc}{\Rc(u) = \Rc(w) \mid \Rc(u) = \Rc(v)}}
    & \geq \ln \frac{\Prb{\Rc}{\Rc(u) = \Rc(v)}}{\Prb{\Rc}{\Rc(u) = \Rc(w)}}
    \nonumber\\& = \ln \frac{1}{\Prb{\Rc}{\Rc(u) = \Rc(w)}} - \ln \frac{1}{\Prb{\Rc}{\Rc(u) = \Rc(v)}}
    \nonumber\\& \approx \frac{\sqrt{d}}{2} - \ln \frac{1}{\Prb{\Rc}{\Rc(u) = \Rc(v)}},\label{naive_cond}
\end{align}
where the last step follows from~(\ref{glsh_lb}), (\ref{glsh_ub}) and the fact that $\|u - w\| \approx \sqrt{2}$.
However this is worse than $\sqrt{d} / 2$ claimed in~(\ref{glsh_cond})
provided that $u$ and $v$ are not too close.
It turns out that~(\ref{naive_cond}) is tight, if we do not assume anything about $\|v - w\|$
(for instance, when $u$, $v$ and $w$ lie on a great circle).
So, we have to ``open the black box'' and derive~(\ref{glsh_cond}) from the first principles.
A high-level idea of the analysis is to observe that, when $\|v - w\| \approx \sqrt{2}$, certain directions
of interest become almost orthogonal, so the corresponding Gaussians are almost independent, which
gives almost independence of the events ``$\Rc(u) = \Rc(w)$'' and ``$\Rc(u) = \Rc(v)$'', which, in turn,
implies
$$
    \ln \frac{1}{\Prb{\Rc}{\Rc(u) = \Rc(w) \mid \Rc(u) = \Rc(v)}}
    \approx \ln \frac{1}{\Prb{\Rc}{\Rc(u) = \Rc(w)}} \approx \frac{\sqrt{d}}{2}
$$
as required. Again, see the full argument in Appendix~\ref{app_glsh}.

\subsection{Implications for ANN}

\label{glsh_ann}

It is illuminating to see what Spherical LSH implies for a random instance (as
defined in the Introduction).  Since all the points lie on a sphere of
radius $cr / \sqrt{2}$, we can plug in
Theorem~\ref{gaussian_lsh_ideal} into Theorem~\ref{lsh_to_nn}, and thus
obtain the exponent
$$
    \rho \leq \frac{1}{2c^2 - 1} + o_c(1). 
$$
Note that we achieve the desired bound~\eqref{eqn:rhoBound} for \emph{random
instances} by using the Spherical LSH directly.

    \section{The data structure}

In this section we describe the new data structure.
First, we show how to achieve success probability $n^{-\rho}$, query time $n^{o_c(1)}$,
and space and preprocessing time $n^{1 + o_c(1)}$, where $\rho = \frac{1}{2c^2 - 1} + o_c(1)$.
Finally, to obtain the final result, one then builds
$O\bigl(n^{\rho}\bigr)$ copies
of the above data structure to amplify the probability of success to $0.99$ (as explained in
Remark \ref{rem:fullDS}).
We analyze the data structure in Section~\ref{apx:analysis}.

\subsection{Overview}
\label{sec_overview}

We start with a high-level overview. Consider a dataset $P_0$ of $n$ points. We can assume that
$r = 1$ by rescaling. 
We may also assume that the
dataset lies in the Euclidean space of dimension $d = \Theta(\log n
\cdot \log \log n)$: one can always reduce the dimension to $d$ by
applying Johnson-Lindenstrauss lemma~\cite{jl-elmhs-84, dg-eptjl-03}
while incurring distortion at most $1 + 1 / (\log \log n)^{\Omega(1)}$
with high probability.

For simplicity, suppose that the entire dataset $P_0$ and a query 
lie on a sphere $\partial B(0, R)$ of radius $R = O_c(1)$.  If
$R \leq c / \sqrt{2}$, we are done: this case corresponds to the
``nicest configuration'' of points and we can apply
Theorem~\ref{lsh_to_nn} equipped with Theorem~\ref{gaussian_lsh_ideal} (see
the discussion in Section~\ref{glsh_ann}).

Now suppose that $R > c / \sqrt{2}$. We split $P_0$ into a number
of disjoint components: $l$ {\em dense} components, termed $C_1$,
$C_2$, \ldots, $C_l$, and one {\em pseudo-random} component, termed
$\widetilde{P}$.  The properties of these components are as follows.
For each dense component $C_i$ we require that $|C_i| \geq \tau n$ and
that $C_i$ can be covered by a spherical cap of radius $(\sqrt{2} -
\eps) R$. Here $\tau, \eps > 0$ are small positive quantities to be
chosen later. The pseudo-random component $\widetilde{P}$ is such that
it contains no more dense components inside.  

\begin{figure}
    \begin{center}
        \includegraphics[page=2,scale=0.91]{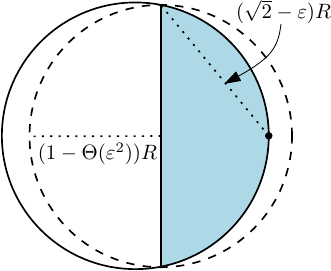}
    \end{center}
    \caption{Covering a spherical cap of radius $(\sqrt{2} - \eps) R$}
    \label{cap_covering_fig}
\end{figure}

We proceed separately for each $C_i$ and $\widetilde{P}$ as follows.
For every dense component $C_i$, we enclose it in a ball $E_i$ of
radius $(1 - \Theta(\eps^2)) R$ (see Figure~\ref{cap_covering_fig}).
For simplicity, let us first ignore the issue that $C_i$ does not
necessarily lie on the boundary $\partial E_i$.  Then, we can just
recurse for the resulting spherical instance with radius $(1 -
\Theta(\eps^2)) R$.  We treat the pseudo-random part $\widetilde{P}$
completely differently.  We sample a partition (hash function)
$\Rc$ of $\partial B(0, R)$ using Theorem~\ref{gaussian_lsh_ideal}.  Then we
partition $\widetilde{P}$ using $\Rc$ and recurse on each non-empty
part. Note that after we have partitioned $\widetilde{P}$, there may
appear new dense clusters in some parts (since it may
become easier to satisfy the minimum size constraint).

During the query procedure, we will recursively
query \emph{each} $C_i$. Furthermore, for the pseudo-random component
$\widetilde{P}$, we
locate the part of $\Rc$ that captures the query point, and
recursively query this part. Overall, there are $(l+1)$ recursive calls.

To analyze our algorithm, we show that we make progress in two ways.
First, for dense clusters we reduce the radius of a sphere by a factor
of $(1 - \Theta(\eps^2))$.  Hence, in $O_c(1 / \eps^2)$ iterations we
must arrive to the case of $R \leq c /\sqrt{2}$, which is easy (as
argued above).  Second, for the pseudo-random component $\widetilde P$,
we argue that most of the points lie at distance $(\sqrt{2} - \eps) R$
from each other. In particular, the ratio of $R$ to a typical
inter-point distance is $\approx 1/\sqrt{2}$, like in a random case
for which Spherical LSH from Theorem~\ref{gaussian_lsh_ideal} is efficient,
as discussed in Section~\ref{glsh_ann}. (This is exactly the reason
why we call $\widetilde{P}$ pseudo-random.) Despite the simplicity
of this intuition, the actual analysis is quite involved: in
particular, this is the place, where we use the three-point
property~(\ref{glsh_cond}) of the Spherical LSH.

It remains to address the issue deferred in the above high-level
description: namely, that a dense component $C_i$ does not generally
lie on $\partial E_i$, but rather can occupy the interior of
$E_i$. We deal with it by partitioning $E_i$ into very thin
annuli of carefully chosen width $\delta$. We then treat each
annulus as a sphere. This discretization of a ball adds to the
complexity of the analysis, although it does not seem
to be fundamental from the conceptual point of view.

Finally, we also show how to obtain \emph{fast} preprocessing, which
turns out to be a non-trivial task, as we discuss in
Section~\ref{fast_prep_sec}.  The main bottleneck is in finding dense
components, for which we show a near-linear time
algorithm.  Roughly, the idea is to restrict ourselves to dense
components with centers in data points: this gives preprocessing time
$n^{2 + o_c(1)}$; we improve it further, to $n^{1 + o_c(1)}$, by sampling
the dataset and searching for dense components in the sample only
(intuitively, this works because we require the dense components to
contain many points).

\subsection{Formal description}

We are now ready to describe the data structure formally. It depends
on the (small positive) parameters $\tau$, $\eps$ and $\delta$, which
we will need to choose carefully later on. 
The pseudocode appears as Figure~\ref{pseudocode}.

\begin{figure}
    \begin{center}
        \includegraphics{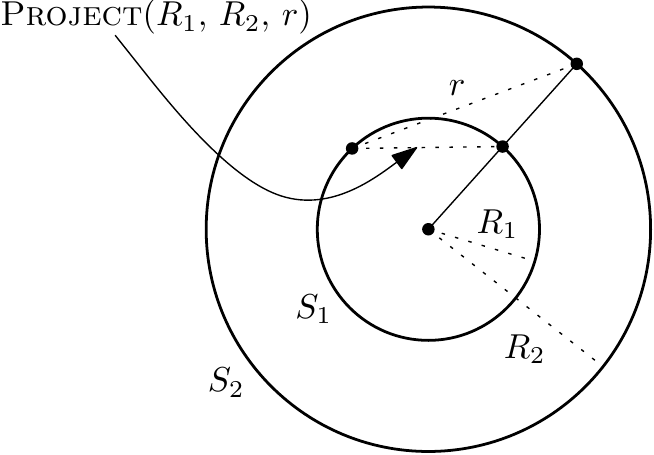}
    \end{center}
    \caption{The definition of \textsc{Project}}
    \label{project_fig}
\end{figure}

\paragraph{Preprocessing.}
Our preprocessing algorithm consists of the following functions:
\begin{itemize}
    \item \textsc{ProcessSphere}($P$, $r_1$, $r_2$, $o$, $R$) builds the data structure for a pointset
    $P$ that lies on a sphere $\partial B(o, R)$,
    assuming we need to solve ANN with distance thresholds $r_1$ and $r_2$. Moreover, we are guaranteed that queries
    will lie on $\partial B(o, R)$, too.
    \item \textsc{ProcessBall}($P$, $r_1$, $r_2$, $o$, $R$) builds the data structure for a dataset 
    $P$ that lies inside the ball $B(o, R)$, assuming we need to solve ANN with distance thresholds $r_1$ and $r_2$.
    Unlike \textsc{ProcessSphere}, here queries can be arbitrary.
    \item \textsc{Process}($P$) builds the data structure for a
      dataset $P$ to solve the general $(1,c)$-ANN;
    \item \textsc{Project}($R_1$, $R_2$, $r$) is an auxiliary
      function computing the following projection. Suppose we have two spheres $S_1$ and $S_2$
    with a common center and radii $R_1$ and $R_2$. Suppose there are points $p_1 \in S_1$ and $p_2 \in S_2$
    with $\|p_1 - p_2\| = r$. \textsc{Project}($R_1$,~$R_2$,~$r$) returns the distance between $p_1$ and the
    point $\widetilde{p_2}$ that lies on $S_1$ and is the closest to $p_2$ (see Figure~\ref{project_fig}).
\end{itemize}

We now elaborate on algorithms in each of the above functions.

\paragraph{\textsc{ProcessSphere}.}  Function \textsc{ProcessSphere}
follows the exposition from Section~\ref{sec_overview}. First, we
consider three base cases. If $r_2 \geq 2R$, then the goal can be
achieved trivially, since any point from $P$ works as an answer for
any valid query.  If $r_1 / r_2 \leq 1 / (2c^2 - 1)$, then
Theorem~\ref{lsh_to_nn} coupled with the hash family from
Theorem~\ref{diim} does the job. Similarly, if $r_2 \geq \sqrt{2} R$,
then we can use the family from Theorem~\ref{gaussian_lsh_ideal} (see
the discussion in Section~\ref{glsh_ann}).  Otherwise, we find
non-trivially smaller balls (of radius $(\sqrt{2} - \eps)R$)
    with centers on $\partial B(o, R)$ that
contain many data points (at least $\tau |P|$).
These balls can be enclosed into balls
(with unconstrained center) of radius $\widetilde{R} \leq (1 -
\Omega(\eps^2)) R$ (proven in Claim~\ref{shrinkage}). For these balls
we invoke \textsc{ProcessBall}.  Then, for the remaining points we
sample a partition of $\partial B(o, R)$ using
Theorem~\ref{gaussian_lsh_ideal}, and recurse on each part.  We note
that in order to apply Theorem~\ref{diim} and
Theorem~\ref{gaussian_lsh_ideal} we need certain conditions on $r_1,
r_2$ and $R$ to hold (we define and verify them in Claim~\ref{stupid_claims}).

\paragraph{\textsc{ProcessBall}.}
First, we consider the following simple base case. If $r_1 + 2R \leq r_2$,
then any point from $B(o, R)$ could serve as a valid answer to any query.

In general, we reduce to the spherical case via a discretization of
the ball $B(o, R)$. First, we round all the distances to $o$ up to a
multiple of $\delta$, which can change distance between any pair of
points by at most $2 \delta$ (by the triangle inequality).  Then, for
every possible distance $\delta i$ from $o$ to a~data point and every
possible distance $\delta j$ from $o$ to a~query (for admissible
integers $i,j$), we build a~separate data structure via
\textsc{ProcessSphere} (we also need to check that $|\delta (i - j)|
\leq r_1 + 2 \delta$ to ensure that the corresponding pair $(i, j)$
does not yield a trivial instance).  We compute the new
distance thresholds $\widetilde{r}_1$ and $\widetilde{r}_2$ for this
data structure as follows.  After rounding, the new thresholds for the
ball instance should be $r_1 + 2 \delta$ and $r_2 - 2 \delta$, since
distances can change by at most $2 \delta$. To compute the final thresholds
(after projecting the query to the sphere of radius $\delta i$), we just
invoke \textsc{Project} (see the definition above).

\paragraph{\textsc{Process}.}  \textsc{Process} reduces the general
case to the ball case. We proceed similarly to \textsc{ProcessSphere},
with a three modifications. First, instead of the family from
Theorem~\ref{gaussian_lsh_ideal}, we use the family from
Theorem~\ref{diim} which is designed for partitioning the whole
$\Rbb^d$ rather than just a sphere.  Second, we seek to find clusters
of radius $2c^2$. Third, we do not need to find the smallest enclosing
ball for $P \cap B(x, 2c^2)$: instead, $B(x, 2c^2)$ itself is enough.

\paragraph{\textsc{Project}.}
This is implemented by a formula (see Figure~\ref{project_fig}).

Overall, the preprocessing creates a decision tree, where the nodes
correspond to procedures {\sc ProcessSphere}, {\sc ProcessBall}, {\sc
  Process}. We refer to the tree nodes correspondingly, using the labels
in the below description of the query algorithm.

Observe that currently the preprocessing is
expensive: a priori it is not even clear how to make it
\emph{polynomial} in $n$ as we need to search over all possible ball centers
$o$. We address this challenge in Section~\ref{fast_prep_sec}.

\paragraph{Query procedure.}
Consider a query point $q \in \Rbb^d$. We run the query on the
decision tree, starting with the root, and applying the following
algorithms depending on the label of the nodes:
\begin{itemize}
\item In \textsc{Process} we first recursively query the ball data structures.
Second, we locate $q$ in $\Rc$, and query the data structure we built for $P \cap \Rc(q)$.
\item In \textsc{ProcessBall}, we first consider the base case, where we just return the stored point if it is
close enough. In general,
we check if $\|q - o\| \leq R + r_1$. If not, we can return.
Otherwise, we round $q$ so that the distance from $o$ to $q$ is a multiple of $\delta$.
Next, we enumerate the distances from $o$ to the potential near neighbor we are looking for,
and query the corresponding {\sc ProcessSphere} children after
projecting $q$ on the sphere with a tentative near neighbor (using, naturally, \textsc{Project}).
\item In \textsc{ProcessSphere}, we proceed exactly the same way as
  \textsc{Process} modulo the base cases, which we handle according to
  Theorem~\ref{lsh_to_nn}.
\end{itemize}

\begin{figure}
\begin{multicols}{2}
{\footnotesize
\begin{algorithmic}
    \Function{Process}{$P$}
        \State $m \gets |P|$
        \While{$\exists\,x \in \Rbb^d: |B(x, 2c^2) \cap P| \geq \tau m$
              }
            \State \Call{ProcessBall}{$P \cap B(x,
                   2c^2)$, $1$, $c$, $x$, $2 c^2$}
            \State $P \gets P \setminus B(x, 2c^2)$
        \EndWhile
        \State sample $\Rc$ according to Theorem~\ref{diim}
        \For{$U \in \Rc$}
            \If{$P \cap U \ne \emptyset$}
                \State \Call{Process}{$P \cap U$}
            \EndIf
        \EndFor
    \EndFunction
    \Function{ProcessBall}{$P$, $r_1$, $r_2$, $o$, $R$}
        \If{$r_1 + 2R \leq r_2$}
            \State store any point from $P$
            \State \Return
        \EndIf
        \State $P \gets \{o + \delta \lceil \frac{\|p - o\|}{\delta}\rceil \cdot \frac{p - o}{\|p - o\|} \mid p \in P\}$
        \For{$i \gets 0 \ldots \lceil \frac{R}{\delta} \rceil$}
            \State $\widetilde{P} \gets \{p \in P \colon \|p - o\| = \delta i\}$
            \If{$\widetilde{P} \ne \emptyset$}
                \For{$j \gets 0 \ldots \lceil
                    \frac{R + r_1 + 2 \delta}{\delta} \rceil$}
                    \If{$\delta |i - j| \leq r_1 + 2 \delta$}
                        \State $\widetilde{r_1} \gets
                        \mbox{\Call{Project}{$\delta i$, $\delta j$,
                        $r_1 + 2 \delta$}}$
                        \State $\widetilde{r_2} \gets
                        \mbox{\Call{Project}{$\delta i$, $\delta j$,
                        $r_2 - 2 \delta$}}$
                        \State
                        \Call{ProcessSphere}{$\widetilde{P}$, $\widetilde{r_1}$, $\widetilde{r_2}$, $o$, $\delta i$}
                    \EndIf
                \EndFor
            \EndIf
        \EndFor
    \EndFunction
    \Function{Project}{$R_1$, $R_2$, $r$}
        \State \Return $\sqrt{R_1 (r^2 - (R_1 - R_2)^2) / R_2}$
    \EndFunction
    \Function{ProcessSphere}{$P$, $r_1$, $r_2$, $o$, $R$}
        \If{$r_2 \geq 2R$}
            \State store any point from $P$
            \State \Return
        \EndIf
        \If{$\frac{r_1}{r_2} \leq \frac{1}{2c^2 - 1}$}
            \State apply Theorem~\ref{lsh_to_nn} with Theorem~\ref{diim} to $P$
            \State \Return
        \EndIf
        \If{$r_2 \geq \sqrt{2} R$}
            \State apply Theorem~\ref{lsh_to_nn} with Theorem~\ref{gaussian_lsh_ideal} to $P$
            \State \Return
        \EndIf
        \State $m \gets |P|$
        \State $\widehat{R} \gets (\sqrt{2} - \eps) R$
        \While{$\exists \, x \in \partial B(o, R): |B(x, \widehat{R}) \cap P| \geq \tau m$
               }
            \State $B(\widetilde{o}, \widetilde{R}) \gets $ the {\sc seb} for
            $P \cap B(x, \widehat{R})$
            \State \Call{ProcessBall}{$P \cap B(x,
                   \widehat{R})$, $r_1$, $r_2$,
                $\widetilde{o}$,
                $\widetilde{R}$}
            \State $P \gets P \setminus B(x,
                   \widehat{R})$ 
        \EndWhile
        \State sample $\Rc$ according to Theorem~\ref{gaussian_lsh_ideal}
        \For{$U \in \Rc$}
            \If{$P \cap U \ne \emptyset$}
                \State \Call{ProcessSphere}{$P \cap U$, $r_1$, $r_2$,
                    $o$, $R$}
            \EndIf
        \EndFor
    \EndFunction
\end{algorithmic}
}
\end{multicols}
\caption{Pseudocode of the data structure (\textsc{seb} stands for \emph{smallest enclosing ball})}
\label{pseudocode}
\end{figure}

\section{Analysis of the data structure}
\label{apx:analysis}

In this section we analyze the above data structure.

\subsection{Overview}

The most interesting part of the analysis is lower bounding the
probability of success: we need to show that it is at least
$n^{-\rho - o_c(1)}$, where $\rho = \frac{1}{2c^2 - 1}$. The challenge is that we need to
analyze a (somewhat adaptive) random process. In particular, we cannot
just use \emph{probability of collision of far points} as is usually done
in the analysis of (data-independent) LSH families. Instead, we use its
\emph{empirical estimate}: namely, the expected number of data points remaining
in the part containing the query $q$. While
this allows to make some good-quality progress, this only lasts for a~%
few iterations of partitioning, until we run into the fact that the
set is only {\em pseudo\/}-random and the deviations from the ``ideal
structure'' begin showing up more prominently (which is the reason
that, after partitioning, we again need to check for densely populated balls).
Furthermore, while computing this empirical estimate, we need to
condition on the fact that the near neighbor is colliding with the
query.

A bit more formally, the proof proceeds in two steps.  First, we show
that whenever we apply a~partition $\Rc$ to a pseudo-random
remainder $P$, the quality we achieve is great: the
exponent we get is $\ln(1 / p_1) / \ln(1 / p_2) \leq \frac{1}{2c^2 - 1} +
o_c(1)$ (see Claim~\ref{decay_rate}). Here $p_1 = \Prb{\Rc}{\Rc(p) =
  \Rc(q)}$ is the probability for the query $q \in \Rbb^d$ and its
near neighbor $p\in P$ to collide under $\Rc$, and
$$
p_2 = \mathrm{E}_\Rc \biggl[\frac{|\Rc(p) \cap P|}{|P|} \biggm| \Rc(p) = \Rc(q)\biggr]$$ is the
(conditioned) empirical estimate of the \emph{fraction of $P$ that collides with $p$}. 
Note that, when computing $p_2$, we
condition on the fact that the query and its near neighbor collide
(i.e., $\Rc(p) = \Rc(q)$). It is exactly this conditioning that
requires the three-point property~(\ref{glsh_cond}) of the Spherical
LSH.  
Furthermore, we use the fact that all the ``dense'' balls have been
carved out, in order to argue that, on average, many points are far
away and so $p_2$ is essentially governed by the collision probability of the Spherical LSH
for distances around $\sqrt{2} \, R$.
In the second step,
we proceed by lower bounding the probability of success via a careful
inductive proof analyzing the corresponding random process
(Claim~\ref{prob_bound}). Along the way, we use the above estimate
crucially. See Section~\ref{success_sec} for details.

The rest of the analysis proves that the data
structure occupies $n^{1 + o_c(1)}$ space and has $n^{o_c(1)}$ query
time (in expectation). While a bit tedious, this is
relatively straightforward.  See Section~\ref{space_sec} for details.
Finally, we highlight that obtaining the near-linear preprocessing
algorithm requires further ideas and in particular utilizes the van
der Corput lemma. See Section~\ref{fast_prep_sec}.

\subsection{Setting parameters}
\label{parameters_sec}
Recall that the dimension is $d =
\Theta( \log n \cdot \log \log n)$.  We set $\eps, \delta, \tau$ as
follows:
\begin{itemize}
    \item $\eps = \frac{1}{\log \log \log n}$;
    \item $\delta = \exp\bigl(-(\log \log \log n)^C\bigr)$;
    \item $\tau = \exp\bigl(-\log^{2/3} n\bigr)$,
\end{itemize}
where $C$ is a sufficiently large positive constant (the concrete value of $C$ is only important
for the proof of Claim~\ref{stupid_claims}).

\subsection{Invariants and auxiliary properties}
We now state and prove several invariants and properties of the data structure that are needed for the subsequent
analysis.

\begin{claim}
    \label{shrinkage}
    In \textsc{ProcessSphere} we have
    $
        \widetilde{R} \leq (1 - \Omega(\eps^2)) R
    $
    (see Figure~\ref{cap_covering_fig}).
\end{claim}
\begin{proof}
    It is enough to show that for $x \in \partial B(o, R)$ there is a ball of radius
    $(1 - \Omega(\eps^2)) R$ that covers $\partial B(o, R) \cap B(x, (\sqrt{2} - \eps) R)$.
    Without loss of generality we can assume that $o = 0$ and $x = (R, 0, \ldots, 0)$.
    Then, $\partial B(0, R) \cap B(x, (\sqrt{2} - \eps) R) =
    \set{u \in \partial B(0, R) \colon u_1 \geq \eta R}$, where $\eta = \Theta(\eps)$.
    At the same time, we can cover $\set{u \in \partial B(0, R) \colon u_1 \geq \eta R}$
    with the ball centered in $(\eta R, 0, \ldots, 0)$ and radius
    $R \sqrt{1 - \eta^2} = (1 - \Omega(\eta^2)) R = (1 - \Omega(\eps^2)) R$. 
\end{proof}

\begin{lemma}
    \label{stupid_claims}
    The following invariants hold.
    \begin{itemize}
    \item At every moment of time we have $\tfrac{r_2}{r_1} \geq c - o_c(1)$, $r_2 \leq c + o_c(1)$
    and $R \leq O_c(1)$;
    \item After checking for the base case in \textsc{ProcessBall}
    and the first base case in \textsc{ProcessSphere} we have $r_2 / R \geq \exp(-O_c(\log \log \log n)^{O(1)})$;
    \item At every moment of the preprocessing or the query procedures,
    the number of calls to \textsc{ProcessBall} in the recursion
    stack is $O_c(\eps^{-O(1)})$;
    \item The expected
    length of any contiguous run of calls to \textsc{Process} or \textsc{ProcessSphere} in the recursion
    stack  is $O(\log n)$ (again, during the preprocessing or the query).
    \end{itemize}
\end{lemma}

The rest of the Section is devoted to proving Lemma~\ref{stupid_claims}. The proofs are quite technical,
and can be omitted on first reading.

Consider the recursion stack at any moment of the preprocessing or the query algorithms.
It has several calls to \textsc{ProcessBall} interleaved with sequences of calls to
\textsc{Process} and \textsc{ProcessSphere}.
Our current goal is to bound the number of calls to \textsc{ProcessBall} that can appear in the recursion
stack at any given moment (we want to bound it by $O_c(1 / \eps^6)$).
First, we prove that this bound holds under the (unrealistic) assumption that
in \textsc{ProcessBall} the rounding of distances has no effect (that is, all the distances of
points to $o$ are already multiples of $\delta$). Second, we prove the bound in full generality
by showing that this rounding introduces only a tiny multiplicative error to $r_1$, $r_2$ and $R$.

\begin{claim}
    \label{vanilla_depth}
    Suppose that the rounding of distances in \textsc{ProcessBall} has
    no effect (i.e., distances from $o$ to all points are multiples of $\delta$).
    Then the number of the calls to \textsc{ProcessBall} in the recursion stack at any given moment
    of time is $O_c(1 / \eps^6)$.
\end{claim}
\begin{proof}

    Let us keep track of two quantities: $\eta = r_2^2 / r_1^2$ and $\xi = r_2^2 / R^2$.
    It is immediate to see that the initial value of $\eta$ is $c^2$, it is non-decreasing
    (it can only change, when we apply \textsc{Project}, which can only increase the ratio
    between $r_2$ and $r_1$), and it is at most $(2c^2 - 1)^2$ (otherwise, the base
    case in \textsc{ProcessSphere} is triggered).
    Similarly, $\xi$ is initially equal to $1 / (4c^2)$ and it can be at most $2$
    (otherwise, the base in \textsc{ProcessSphere} is triggered).
    Unlike $\eta$, the value of $\xi$ \emph{can} decrease.

    Suppose that in \textsc{ProcessBall} we call \textsc{ProcessSphere} for some $R_1 = \delta i$
    and $R_2 = \delta j$ with $|R_1 - R_2| = \Delta R$.
    Suppose that $\widetilde{\eta}$ is the new value of $\eta$ after this call.
    We have
    \begin{multline}
        \label{eta_evolution}
        \frac{\widetilde{\eta}}{\eta} = \frac{\widetilde{r_2}^2 / \widetilde{r_1}^2}{r_2^2 / r_1^2}
        = \frac{\mbox{\textsc{Project}($R_1$, $R_2$, $r_2$)}^2 /
        \mbox{\textsc{Project}($R_1$, $R_2$, $r_1$)}^2}{r_2^2 / r_1^2}
        = \frac{(r_2^2 - \Delta R^2) / (r_1^2 - \Delta R^2)}{r_2^2 / r_1^2}
        \\ = \frac{1 - \frac{\Delta R^2}{r_2^2}}{1 - \frac{\Delta R^2}{r_1^2}}
        = \frac{1 - \frac{\Delta R^2}{r_2^2}}{1 - \eta \cdot \frac{\Delta R^2}{r_2^2}}
        = 1 + \Omega_c \Big(\frac{\Delta R^2}{r_2^2}\Big),
    \end{multline}
    where the third step follows from the formula for \textsc{Project} and the last
    step follows from the fact that $\eta \geq c^2 > 1$.

    Let us call an invocation of \textsc{ProcessSphere} within \textsc{ProcessBall}
    \emph{$\lambda$-shrinking} for some $\lambda > 0$,
    if $\Delta R^2 / r_2^2 \geq \lambda$. From~(\ref{eta_evolution}), the fact that
    $\eta \in [c^2; (2c^2 - 1)^2]$ and that $\eta$ is non-decreasing, we conclude that
    there can be at most $O_c(1 / \lambda)$ $\lambda$-shrinking invocations of \textsc{ProcessSphere}
    in the recursive stack at any given moment of time.

    Now let us see how $\xi$ evolves. Suppose that within some \textsc{ProcessBall}
    we call \textsc{ProcessSphere} with some $R_1$ and $R_2$. Then, in \textsc{ProcessSphere}
    we call \textsc{ProcessSphere} recursively several times without any change in $r_1$, $r_2$ and
    $R$, and finally we call \textsc{ProcessBall} again.
    Denote $\widetilde{\xi} = \frac{\widetilde{r_2}^2}{\widetilde{R}^2}$
    the new value of $\xi$ after this call of \textsc{ProcessBall}. We have
    \begin{multline}
        \label{xi_evolution}
        \frac{\widetilde{\xi}}{\xi}
        =
        \frac{\widetilde{r_2}^2 / \widetilde{R}^2}{r_2^2 / R^2}
        \geq (1 + \Omega(\eps^2))
        \frac{\widetilde{r_2}^2 / R_1^2}{r_2^2 / R^2}
        = (1 + \Omega(\eps^2)) \frac{(r_2^2 - \Delta R^2) / (R_1 R_2)}{r_2^2 / R^2}
        \\ = (1 + \Omega(\eps^2)) \Big( 1 - \frac{\Delta R^2}{r_2^2} \Big) \frac{R^2}{R_1 R_2}
        \geq (1 + \Omega(\eps^2)) \Big( 1 - \frac{\Delta R^2}{r_2^2} \Big) \frac{1}{1 + \frac{\Delta R}{R}},
    \end{multline}
    where the second step follows from Claim~\ref{shrinkage},
    the third step follows from the formula for \textsc{Project},
    the fifth step follows from the fact that $R_1 \leq R$ and $R_2 \leq R_1 + \Delta R \leq R + \Delta R$.

    Denote $\lambda^* = \eps^4 / C$ for a sufficiently large constant $C > 0$.
    If the call to \textsc{ProcessSphere} within \textsc{ProcessBall}
    is not $\lambda^*$-shrinking (that is, $\Delta R / r_2 \leq \eps^2 / \sqrt{C})$, then
    since
    $$
        \frac{\Delta R}{R} = \frac{\Delta R}{r_2} \cdot \sqrt{\xi}
        \leq \frac{\sqrt{2} \cdot \Delta R}{r_2} \leq \sqrt{\frac{2}{C}} \cdot \eps^2,
    $$
    where we use $\xi \leq 2$,
    from~(\ref{xi_evolution}) we have that $\widetilde{\xi} / \xi = 1 + \Omega(\eps^2)$
    (provided that $C$ is sufficiently large).
    On the other hand, if the call is $\lambda^*$-shrinking, then since
    $$
        \frac{\Delta R^2}{r_2^2} \leq \frac{r_1^2}{r_2^2} \leq \frac{1}{c^2},
    $$
    and
    $$
        \frac{\Delta R}{R} \leq \frac{r_1}{R} = \frac{r_1}{r_2} \cdot \frac{r_2}{R}
        \leq \frac{\sqrt{2}}{c},
    $$
    we have from~(\ref{xi_evolution})
    $$
        \frac{\widetilde{\xi}}{\xi} \geq \Big(1 - \frac{1}{c^2}\Big) \frac{1}{1 + \sqrt{2} / c} 
        = \Omega_c(1),
    $$
    since $c > 1$.
    That being said, non-$\lambda^*$-shrinking calls increase $\xi$
    non-trivially (by at least $(1 + \Omega(\eps^2))$), while $\lambda^*$-shrinking calls decrease $\xi$
    by not too much, and by the above discussion, there are at most $O_c(1 / \lambda^*) = O_c(1 / \eps^4)$
    of them.

    More formally, suppose we have $A$ calls to \textsc{ProcessSphere} that are $\lambda^*$-shrinking
    and $B$ calls that are not $\lambda^*$-shrinking.
    We have that $A = O_c(1 / \lambda^*) = O_c(1 / \eps^4)$.
    On the other hand, since every $\lambda^*$-shrinking call multiplies $\xi$
    by at least $\Omega_c(1)$, the initial value of $\xi$ is $1 / (4c^2)$,
    and the final value of $\xi$ is at most $2$, we have
    $$
        (1 + \Omega(\eps^2))^B \leq \exp(O_c(A)),
    $$
    thus, $B = O_c(1 / \eps^6)$.
    Overall, we have at most $A + B \leq O_c(1 / \eps^6)$ invocations of \textsc{ProcessBall}
    in the recursion stack at any moment of time.
\end{proof}

\begin{claim}
    \label{vanilla_magnitude}
    Suppose that the rounding of distances in \textsc{ProcessBall} has
    no effect (i.e., distances from $o$ to all points are multiples of $\delta$).
    At any moment of time, in \textsc{ProcessBall} and \textsc{ProcessSphere}
    outside of the base cases
    one has $r_1, r_2, R \geq \exp(-O_c(1 / \eps^{10}))$.
\end{claim}
\begin{proof}
    By the proof of Claim~\ref{vanilla_depth}, $r_2 / r_1$ is $\Theta_c(1)$ and
    $r_2 / R \in [\exp(-O_c(1 / \eps^4)); \sqrt{2}]$.

    After calling \textsc{ProcessSphere} within \textsc{ProcessBall} (and vice versa)
    the new value of $R$ is
    at least $\Omega(r_1)$, since otherwise the first base case in \textsc{ProcessSphere} (or
    the base case of \textsc{ProcessBall}) will be triggered.

    So, overall we have $O_c(1 / \eps^6)$ calls to \textsc{ProcessBall} in the recursion stack
    each of which can decrease $r_1, r_2, R$ by a factor of at most $\exp(O_c(1 / \eps^4))$. Hence the claim.
\end{proof}

Since $\eps = 1 / \log \log \log n$, we can choose $C$ in the definition of $\delta$ (see Section~\ref{parameters_sec}) so that, by Claim~\ref{vanilla_magnitude},
rounding distance to multiplies of $\delta$ gives only a small \emph{multiplicative}
change to $r_1, r_2, R$ that accumulates to $1 + o_c(1)$ over $O_c(1 / \eps^6)$ iterations.

This way, we obtain all the items in Claim~\ref{stupid_claims}, except the last one
(by staring at the above proofs and taking the previous paragraph into account).

Let us show the last item for \textsc{Process} (for \textsc{ProcessSphere} the proof is the same
verbatim).
Let us look at any data point $p \in P_0$. Suppose that $p$ ends up in a pseudo-random remainder.
Then, there are only $\tau n$ points in the $2c^2$-neighborhood of $p$.
When we sample $\Rc$, the number of points outside this neighborhood multiplies by a constant
stricly smaller than one (in expectation). Thus, overall, the number of points in $P$ multiplies by a constant
smaller than one on average every call to \textsc{Process}. It means that in $O(\log n)$ calls with high probability
either $p$ will be captured by a dense cluster or it will be the only remaining point.

\subsection{Probability of success}
\label{success_sec}
We now lower bound the probability of success for a query $q \in \Rbb^d$
for which there exists $p \in P_0$ with $\|q - p\| \leq 1$.  We
perform this analysis in two steps. First, we upper bound
$\Exp{\Rc}{|P \cap \Rc(q)| \mid \Rc(p) = \Rc(q)}$ in \textsc{Process}
and \textsc{ProcessSphere} provided that $p \in P$ after removing
dense clusters. This upper bound formalizes the intuition that after
removing dense clusters the remaining pseudo-random instance becomes
easy to partition using Spherical LSH.  While upper bounding this expectation for
\textsc{ProcessSphere}, we crucially rely on the estimate~\eqref{glsh_cond} for triples of points from
Theorem~\ref{gaussian_lsh_ideal} (see also the remark after the proof
of Claim~\ref{decay_rate}).  Second, we use this estimate to lower
bound the overall probability of success by analyzing the
corresponding random process.

\begin{figure}
    \begin{center}
        \includegraphics{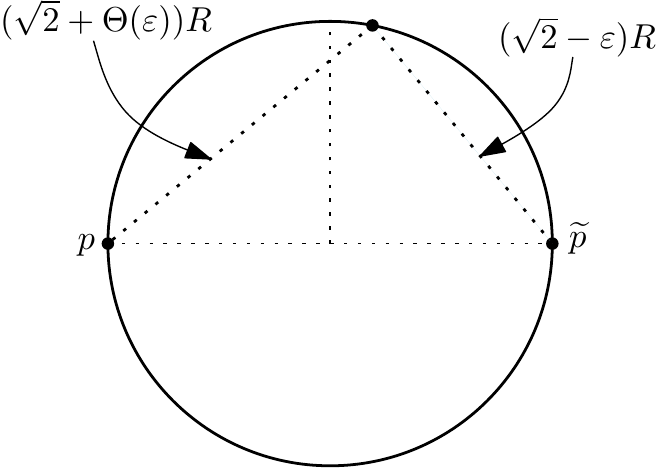}
    \end{center}
    \caption{For the proof of Claim~\ref{decay_rate}: distances to points in $U$}
    \label{opposite_fig}
\end{figure}

\begin{claim}
    \label{decay_rate}
    Suppose that in \textsc{Process} or \textsc{ProcessSphere}, we have
    $p \in P$ after removing all dense balls.
    Let $q \in \Rbb^d$ be such that $\|p - q\| \leq 1$ for \textsc{Process} or $\|p - q\| \leq r_1$
    and $q \in \partial B(o, R)$
    for \textsc{ProcessSphere}.
    Then, after sampling $\Rc$, we have
    \begin{equation}
        \label{decay_expr}
        \frac{\ln(1 / p_1)}{\ln(1 / p_2)} \leq \frac{1}{2c^2 - 1} + o_c(1),
    \end{equation}
    where $p_1 = \Prb{\Rc}{\Rc(p) = \Rc(q)}$,
    and $p_2 = \Exp{\Rc}{\tfrac{|P \cap \Rc(q)|}{m} \mid \Rc(p) = \Rc(q)}$,
    where $m$ is the size of $P$ at the beginning of \textsc{Process} or
    \textsc{ProcessSphere}.
\end{claim}
\begin{proof}
    Let us first analyze \textsc{Process}, for which the analysis is somewhat simpler.
    In this case, by Theorem~\ref{diim}, we have
    \begin{equation}
        \label{decay_0}
        \ln(1 / p_1) = \ln \frac{1}{\Prb{\Rc}{\Rc(p) = \Rc(q)}} = (1 + o(1)) \sqrt{d}.
    \end{equation}
    On the other hand,
    \begin{align}
        p_2 = \frac{\Exp{\Rc}{|P \cap \Rc(q)| \mid \Rc(p) = \Rc(q)}}{m} & \leq \tau
         + \frac{\Exp{\Rc}{|(P \cap \Rc(q)) \setminus B(q, 2c^2)| \mid \Rc(p) = \Rc(q)}}{m}
        \nonumber\\& \leq \tau + \sup_{p' \in P \setminus B(q, 2c^2)} \Prb{\Rc}{\Rc(p') = \Rc(q) \mid \Rc(p) = \Rc(q)}
        \nonumber\\& \leq \tau + \frac{\sup_{p' \in P \setminus B(q, 2c^2)} \Prb{\Rc}{\Rc(p') = \Rc(q)}}{\Prb{\Rc}{\Rc(p) = \Rc(q)}}
        \nonumber\\& \leq \tau + e^{-(2c^2 - 1) \sqrt{d} \cdot (1 + o_c(1))}
        \nonumber\\& \leq e^{-(2c^2 - 1) \sqrt{d} \cdot (1 + o_c(1))}, \label{decay_1}
    \end{align}
    where the first step follows from the fact that after removing the dense clusters we have
    $|\Rc \cap B(q, 2c^2)| < \tau m$, the second step follows from the linearity of expectation,
    the fourth step follows from Theorem~\ref{diim} and the last step uses
    that $d = \Theta(\log n \log \log n)$ and $\tau = \exp(- \log^{2/3} n)$.
    Now, combining~(\ref{decay_0}) and~(\ref{decay_1}), we get
    $$
        \frac{\ln(1 / p_1)}{\ln(1 / p_2)} \leq \frac{1}{2c^2 - 1} + o_c(1)
    $$
    as desired.

    Now let us analyze \textsc{ProcessSphere}.
    By Claim~\ref{stupid_claims} and the fact that $r_2 / r_1 \leq O_c(1)$
    (otherwise, we would have triggered the second base case in \textsc{ProcessSphere}), we have $r_1 / R \geq \exp(-O_c(\log \log \log n)^{O(1)})$,
    and $d = \Theta(\log n \log \log n)$,
    so we are in position to apply Theorem~\ref{gaussian_lsh_ideal}.
    We have
    \begin{align}
        \ln(1 / p_1) & \leq \frac{(r_1 / R)^2}{4 - (r_1 / R)^2}\cdot \frac{\sqrt{d}}{2} \cdot (1 + o(1))
        \nonumber\\ & \leq \frac{1}{2c^2 - 1} \cdot \frac{\sqrt{d}}{2} \cdot (1 + o_c(1)),\label{decay_2}
    \end{align}
    where the second step follows from the estimate
    $$
        \frac{r_1}{R} = \frac{r_1}{r_2} \cdot \frac{r_2}{R} \leq \frac{\sqrt{2}}{c} \cdot (1 + o_c(1)),
    $$
    where the second step is due to the assumptions $r_2 / r_1 \geq c - o_c(1)$ (which is true
    by Claim~\ref{stupid_claims}) and
    $r_2 \leq \sqrt{2} R$ (if the latter was not the case, we would have triggered the third base case
    in \textsc{ProcessSphere}).
    Let $\widetilde{p}$ and $\widetilde{q}$ be reflections of $p$ and $q$ respectively with respect
    to $o$. Define
    \begin{align*}
         U & = B(q, (\sqrt{2} - \eps)R) \cup B(\widetilde{q}, (\sqrt{2} - \eps)R)
        \\ & \cup B(p, (\sqrt{2} - \eps)R) \cup B(\widetilde{p}, (\sqrt{2} - \eps)R).
    \end{align*}
    
    Then,
    \begin{align}
        p_2 = \frac{\Exp{\Rc}{|P \cap \Rc(q)| \mid \Rc(p) = \Rc(q)}}{m} & \leq 4 \tau
         + \frac{\Exp{\Rc}{|(P \cap \Rc(q)) \setminus U| \mid \Rc(p) = \Rc(q)}}{m}
        \nonumber\\& \leq 4 \tau + \sup_{p' \in P \setminus U} \Prb{\Rc}{\Rc(p') = \Rc(q) \mid \Rc(p) = \Rc(q)}
        \nonumber\\& \leq 4 \tau + e^{-\frac{\sqrt{d}}{2} \cdot (1 + o(1))}
        \nonumber\\& \leq e^{-\frac{\sqrt{d}}{2} \cdot (1 + o(1))},\label{decay_3}
    \end{align}
    where the first step follows from the definition of $U$ and that we have all the dense clusters removed,
    the second step follows from the linearity of expectation,
    the third step follows from Theorem~\ref{gaussian_lsh_ideal} (namely,~(\ref{glsh_cond}))
    and the fact that all the points from $P \setminus U$
    are within $(\sqrt{2} \pm \Theta(\eps)) R$ from
    \emph{both} $p$ and $q$ (see Figure~\ref{opposite_fig}),
    and the fourth step uses the values for $d$ and $\tau$.
    Overall, combining~(\ref{decay_2}) and~(\ref{decay_3}), we get
    $$
        \frac{\ln(1 / p_1)}{\ln(1 / p_2)} \leq \frac{1}{2c^2 - 1} + o_c(1).
    $$
\end{proof}

\paragraph{Remark:} If instead of~(\ref{glsh_cond}) we used~(\ref{naive_cond}), then the best we could hope for 
would be
$$
    \frac{\ln(1 / p_1)}{\ln(1 / p_2)} \leq \frac{1}{2c^2 - 2} + o_c(1),
$$
which is much worse than~(\ref{decay_expr}), if $c$ is close to $1$.

\begin{claim}
    \label{prob_bound}
    Suppose that $p \in P_0$ and $q \in \Rbb^d$ with $\|p - q\| \leq 1$.
    Then, the probability of success of the data structure is at least
    $$
        n^{- \frac{1}{2c^2 - 1} - o_c(1)}.
    $$
\end{claim}
\begin{proof}
    Let us prove by induction that any query of a data structure built by
    \textsc{Process}, \textsc{ProcessBall} and \textsc{ProcessSphere}
    with $p \in P$ has probability of success at least $|P|^{-\rho} \cdot n^{-\alpha}$,
    where $\rho \leq 1 / (2c^2 - 1) + o_c(1)$ and $\alpha = o_c(1)$.
    First, we prove this for \textsc{Process} assuming that the same bound holds
    for \textsc{ProcessBall}.
    Then, we argue that for \textsc{ProcessBall} and \textsc{ProcessSphere} essentially the 
    same argument works as well.

    Let us denote $f(m)$ a lower bound on the probability of success for \textsc{Process}
    when $|P| \leq m$ and denote $p_1$ and $p_2$ the quantities from Claim~\ref{decay_rate}
    introduced for \textsc{Process}.

    Let us lower bound $f(m)$ by induction.
    If $p$ belongs to one of the dense clusters,
    then $f(m) \geq m^{-\rho} \cdot n^{-\alpha}$ by the assumption for
    \textsc{ProcessBall}. If $p$ does not belong to any dense cluster, we have
    \begin{align*}
        f(m) & \geq \Exp{\Rc}{f(|P \cap \Rc(q)|) \cdot \mathbf{1}_{\Rc(p) = \Rc(q)}}
        \\ & = \Prb{\Rc}{\Rc(p) = \Rc(q)} \cdot \Exp{\Rc}{f(|P \cap \Rc(q)|) \mid \Rc(p) = \Rc(q)}
        \\ & \geq p_1 \cdot \inf_{\begin{smallmatrix}\supp X \subseteq [m]:\\ \Exp{}{X} \leq p_2 \cdot m\end{smallmatrix}} \Exp{}{f(X)}
        \\ & \geq p_1 \cdot \inf_{\begin{smallmatrix}\supp X \subseteq [m]:\\ \Exp{}{X} \leq p_2 \cdot m\end{smallmatrix}} \Prb{}{X < m} \cdot \Exp{}{f(X) \mid X < m}
        \\& \geq p_1 (1 - p_2) \cdot \inf_{\begin{smallmatrix}\supp Y \subseteq [m-1]:\\ \Exp{}{Y} \leq p_2 \cdot m\end{smallmatrix}} \Exp{}{f(Y)}
        \\& \geq p_1 (1 - p_2) n^{-\alpha} \cdot \inf_{\begin{smallmatrix}\supp Y \subseteq [m-1]:\\
            \Exp{}{Y} \leq p_2 \cdot m\end{smallmatrix}} \Exp{}{Y^{-\rho}}
        \\& \geq p_1 (1 - p_2) n^{-\alpha} \cdot \inf_{\begin{smallmatrix}\supp Y \subseteq [m-1]:\\
            \Exp{}{Y} \leq p_2 \cdot m\end{smallmatrix}} \Exp{}{Y}^{-\rho}
        \\& \geq p_1 (1 - p_2) p_2^{-\rho}m^{-\rho} n^{-\alpha},
    \end{align*}
    where the third step is due to the definition of $p_1$ and $p_2$,
    the fifth step is due to Markov's inequality, the sixth
    step is by the induction hypothesis and the seventh step is due to Jensen's inequality.
    For the induction to go through we must have
    \begin{equation}
        p_1 (1 - p_2) p_2^{-\rho} \geq 1, 
    \end{equation}
    so by Claim~\ref{decay_rate} and the fact that $p_2 \leq 1 - \Omega(1)$
    (which follows from Theorem~\ref{diim}),
    we can set $\rho \leq \frac{1}{2c^2 - 1} + o_c(1)$.

    For \textsc{ProcessBall} and \textsc{ProcessSphere} we can perform a similar induction
    with two modifications. First, we need to verify the bound for the base cases in
    \textsc{ProcessSphere} (in particular, this will determine the value of $\alpha$).
    Second, since \textsc{ProcessBall} and \textsc{ProcessSphere}
    depend on each other, we need to carry on the ``outer'' induction on the number
    of calls to \textsc{ProcessBall} in the recursion stack.
    The latter issue is easy to address, since by Claim~\ref{stupid_claims} the maximum
    number of calls to \textsc{ProcessBall} in the recursion stack is bounded,
    and from the above estimate it is immediate to see that there is no ``error'' that would accumulate.
    Thus, it is only left to look at the base cases of \textsc{ProcessSphere}.
    The first base case is trivial. For the remaining cases we use Theorem~\ref{lsh_to_nn}.
    For the second base case we use the
    Theorem~\ref{diim}: we are in position to apply it, since by Claim~\ref{stupid_claims}
    $r_2 \leq c + o_c(1)$. As for the third case, since by Claim~\ref{stupid_claims}
    and the fact that for the third case $r_2 / r_1 = O_c(1)$ one has that
    $r_1$ is not too small compared to $R$, we are in position to apply Theorem~\ref{gaussian_lsh_ideal}.
    Since by Claim~\ref{stupid_claims} $r_2 / r_1 \geq c - o_c(1)$ and by the assumption
    $r_2 \geq \sqrt{2} R$, applying Theorem~\ref{gaussian_lsh_ideal}, we get the required bound
    on the probability of success.
\end{proof}

\subsection{Space and query time}
\label{space_sec}
In this section we show that the expected space the data structure occupies is $n^{1 + o_c(1)}$
and the expected query time
is $n^{o_c(1)}$.

\begin{claim}
    \label{space_bound}
    The overall expected space the data structure occupies is $n^{1 + o_c(1)}$.
\end{claim}
\begin{proof}
    Every particular point $p \in P_0$ can participate in several branches of recursion during the
    preprocessing: the reason for this is \textsc{ProcessBall}, where every data point participates
    in many sphere data structures.
    But observe that by Claim~\ref{stupid_claims} every point can end up in at most
    \begin{equation}
        \label{num_branches}
        O_c\Big(\frac{1}{\delta}\Big)^{O_c(\eps^{-O(1)})} = n^{o_c(1)}
    \end{equation}
    branches, since there are at most $O_c(\eps^{-O(1)})$ calls to \textsc{ProcessBall}
    in every branch, and each such call introduces branching factor of at most
    $O((R + r_1 + 2 \delta) / \delta) = O_c(1 / \delta)$.

    Next, for every point $p \in P_0$ and for every branch it participates in,
    one can see that by Claim~\ref{stupid_claims} the total expected number of calls in the stack is
    at most
    \begin{equation}
        \label{branch_length}
        O_c\Big(\frac{\log n}{\eps^{O(1)}}\Big) = n^{o_c(1)},
    \end{equation}
    since the number of \textsc{ProcessBall}'s is at most $O_c(\eps^{-O(1)})$
    they separate the runs of \textsc{Process} and \textsc{ProcessSphere}, each of which
    is of length $O(\log n)$ in expectation.

    Since every partition $\Rc$ sampled in \textsc{Process} or \textsc{ProcessSphere}
    takes $n^{o_c(1)}$ space by Theorem~\ref{diim} and Theorem~\ref{gaussian_lsh_ideal},
    we get, combining~(\ref{num_branches}) and~(\ref{branch_length}),
    that the space consumption of partitions and hash tables per point is
    $n^{o_c(1)}$.
    Also, the base cases in \textsc{ProcessSphere} are cheap too: indeed,
    from Theorem~\ref{lsh_to_nn} (coupled with Theorem~\ref{gaussian_lsh_ideal} and Theorem~\ref{diim})
    we see that 
    the space consumption for the base cases is $n^{o_c(1)}$ per point per branch.

    Thus, the overall bound $n^{1 + o_c(1)}$ follows.
\end{proof}

\begin{claim}
    For every query $q \in \Rbb^d$, the expected query time is at
    most $n^{o_c(1)}$.
\end{claim}
\begin{proof}
    Consider a recursion tree for a particular query $q \in \Rbb^d$.

    First, observe that each sequence of calls to \textsc{Process} or \textsc{ProcessSphere}
    can spawn at most $O(\tau^{-1} \cdot \log n)$ calls to \textsc{ProcessBall},
    since every such call multiplies the number of remaining points by a factor
    of at most $(1 - \tau)$.
    At the same time by Claim~\ref{stupid_claims},
    each such sequence is of size $O(\log n)$ in expectation.

    Since by Claim~\ref{stupid_claims} in the recursion stack there can be at most $O_c(\eps^{-O(1)})$ calls
    to \textsc{ProcessBall}, which induces the branching factor
    of at most $O_c(1 / \delta)$, overall, the expected number of nodes in the tree is
    at most
    $$
        \Big(\frac{\log n}{\delta \tau}\Big)^{O_c(\eps^{-O(1)})} = n^{o_c(1)}.
    $$
    In every node we need to do one point location in a partition, which by Theorem~\ref{diim} and
    Theorem~\ref{gaussian_lsh_ideal} can be done in time $n^{o_c(1)}$, and then for the
    base cases we have the expected query time $n^{o_c(1)}$ (by Theorem~\ref{lsh_to_nn}).
    Thus, the overall expected query time is $n^{o_c(1)}$.
\end{proof}

\section{Fast preprocessing}
\label{fast_prep_sec}

A priori, it is not clear how to implement the preprocessing in polynomial time, let alone near-linear.
We will first show how to get preprocessing time to $n^{2 + o_c(1)}$ and then reduce it to $n^{1 + o_c(1)}$.

\subsection{Near-quadratic time}
To get preprocessing time $n^{2 + o_c(1)}$ we need to observe that during the clustering step in \textsc{Process}
and \textsc{ProcessSphere} we may look only for balls with centers being points from $P$.

For \textsc{Process} it is easy to see that we can find balls of radius $4c^2$ with centers being points from
$P$. Then, the proofs of Claim~\ref{decay_rate} and, as a result, Claim~\ref{prob_bound} go through
(we use that, as a result of such a preprocessing, there are no dense clusters
of radius $2c^2$ with \emph{arbitrary} centers). Also, we need to make sure that Claim~\ref{stupid_claims}
is still true, despite we start with clusters of radius $4c^2$ instead of $2c^2$, but that is straightforward.

For \textsc{ProcessSphere} the situation is slightly more delicate, since we can not afford to lose a factor
of two in the radius here.
We build upon the following Lemma.
\begin{proposition}[van der Corput Lemma]
    \label{vdcl}
    For any $v^*, v_1, v_2, \ldots, v_n \in S^{d-1}$ one has
    $$
        \sum_{i, j} \langle v_i, v_j\rangle \geq \Big|\sum_{i} \langle v^*, v_i\rangle\Big|^2.
    $$
\end{proposition}
\begin{proof}
    We have
    $$
        \Big|\sum_{i} \langle v^*, v_i\rangle\Big|^2 = 
        \Big|\big\langle v^*, \sum_i v_i\big\rangle\Big|^2 \leq 
        \|v^*\|^2 \cdot \big\|\sum_i v_i\big\|^2 = \big\|\sum_i v_i\big\|^2 = \sum_{i,j} \langle v_i, v_j\rangle,
    $$
    where
    the second step is an application of the Cauchy-Schwartz inequality.
\end{proof}
The following Claim is the main estimate we use to analyze the variant of \textsc{ProcessSphere}, where we are looking
only for clusters with centers in data points.
Informally, we prove that if a non-trivially small
spherical cap covers $n$ points, then there is a non-trivially small 
cap \emph{centered in one of the points} that covers a substantial fraction of points.
\begin{claim}
    Fix $\eps > 0$. Suppose that $U \subset S^{d-1}$ with $|U| = n$. Suppose that there exists $u^* \in S^{d-1}$ such that
    $\|u^* - u\| \leq \sqrt{2} - \eps$ for every $u \in U$. Then, there exists $u_0 \in U$ such that
    $$
        \left|\set{u \in U \colon \|u - u_0\| \leq \sqrt{2} - \Omega(\eps^2)}\right| \geq \Omega(\eps^2 n).
    $$
\end{claim}
\begin{proof}
    First, observe that $\|u^* - u\| \leq \sqrt{2} - \eps$ iff $\langle u^*, u\rangle \geq \Omega(\eps)$.
    By van der Corput Lemma (Proposition~\ref{vdcl}),
    $$
        \sum_{u, v \in U} \langle u, v \rangle \geq \left|\sum_{u \in U} \langle u^*, u\rangle\right|^2
        \geq \Omega(\eps^2 n^2).
    $$
    Thus, there exists $u_0 \in U$ such that
    $$
        \sum_{u \in U} \langle u_0, u\rangle \geq \Omega(\eps^2 n).
    $$
    This implies
    $$
        \left|\set{u \in U \mid \langle u_0, u \rangle \geq \Omega(\eps^2)}\right| \geq \Omega(\eps^2 n),
    $$
    which is equivalent to
    $$
        \left|\set{u \in U \mid \|u - u_0\| \leq \sqrt{2} - \Omega(\eps^2)}\right| \geq \Omega(\eps^2 n).
    $$
\end{proof}

It means that if in \textsc{ProcessSphere}
we search for clusters of radius $\sqrt{2} - \Omega(\eps^2)$ centered in data points
that cover at least $\Theta(\eps^2 \tau m)$ points, then after we remove all of them,
we are sure that there are no clusters of radius $\sqrt{2} - \eps$ with \emph{arbitrary}
centers that cover at least $\tau m$ points, so Claims~\ref{decay_rate} and~\ref{prob_bound} go through.
The proof of Claim~\ref{stupid_claims} needs to be adjusted accordingly, but we claim that by setting
$C$ in the definition of $\delta$ large enough, the proof of Claim~\ref{stupid_claims}
still goes through.

It is easy to see that by reducing the time of each clustering step to
near-quadratic, we reduce the total preprocessing time to $n^{2 +
  o_c(1)}$. This follows from the proof of Claim~\ref{space_bound}
(intuitively, each point participates in $n^{o_c(1)}$ instances of the
clustering subroutine) and from the fact that it takes time
$n^{o_c(1)}$ to sample a partition $\Rc$ in \textsc{Process} and
\textsc{ProcessSphere}.

\subsection{Near-linear time}

To get $n^{1 + o_c(1)}$ preprocessing time, we just sample the dataset
and compute dense balls in the sample.  Indeed, since we care about
the clusters with at least $\eps^2 \tau n = n^{1 - o_c(1)}$ data
points, we can sample $n^{o_c(1)}$ points from the dataset and find
dense clusters for the sample. Then, using the fact that the VC-dimension
for balls in $\Rbb^d$ is $O(d)$, we can argue that this sample is
accurate enough with probability at least $1 - n^{-10}$. Then, taking
the union bound over all clustering steps, we are done.

    \section{Acknowledgments}

We thank Piotr Indyk and Sepideh Mahabadi for insightful discussions
about the problem and for reading early drafts of this write-up.
In particular, discussions with them led us to the fast preprocessing algorithm.

    \singlespacing
             {\small
    \bibliographystyle{alpha}
    \bibliography{desk}
             }
    \onehalfspacing
    \appendix
    \section{Analysis of Spherical LSH}
\label{app_glsh}

In this section we prove Theorem~\ref{gaussian_lsh_ideal}. We will use
the following basic estimate repeatedly.

\begin{lemma}[e.g., \cite{kms-agcsp-98}]
    \label{gaussians}
    For every $t > 0$
    $$
        \frac{1}{\sqrt{2 \pi}} \cdot \left(\frac{1}{t} - \frac{1}{t^3}\right)
        \cdot e^{-t^2 / 2} \leq \mathrm{Pr}_{X \sim N(0, 1)}[X \geq t]
        \leq \frac{1}{\sqrt{2 \pi}} \cdot \frac{1}{t} \cdot e^{-t^2 / 2}.
    $$
\end{lemma}

\subsection{Collision probability for a pair}
\begin{figure}
    \begin{center}
    \begin{subfigure}{0.47\textwidth}
    \includegraphics[width=\textwidth]{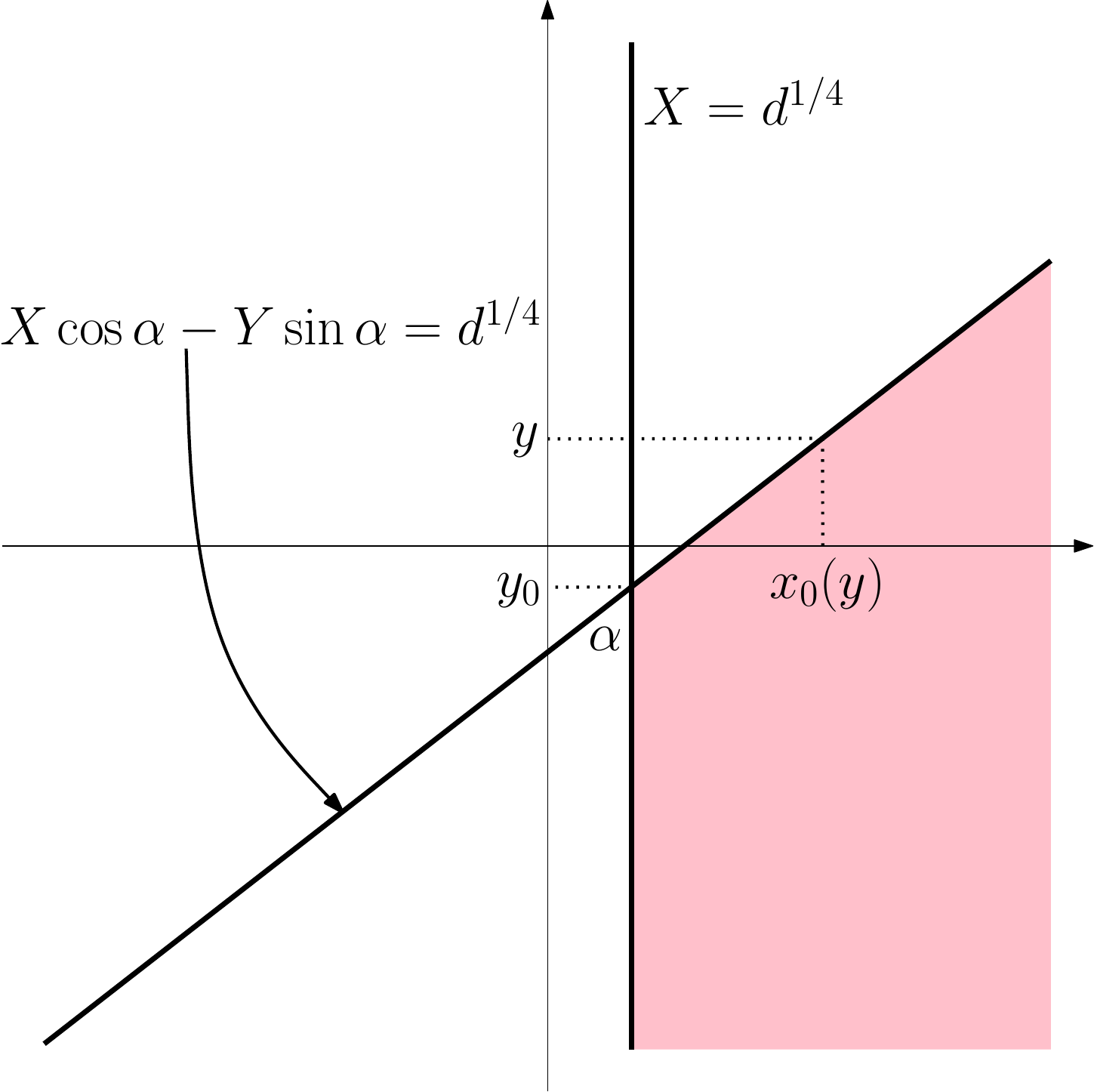}
        \caption{The case $\alpha < \pi / 2$}
    \end{subfigure}
    \quad
    \begin{subfigure}{0.47\textwidth}
    \includegraphics[width=\textwidth]{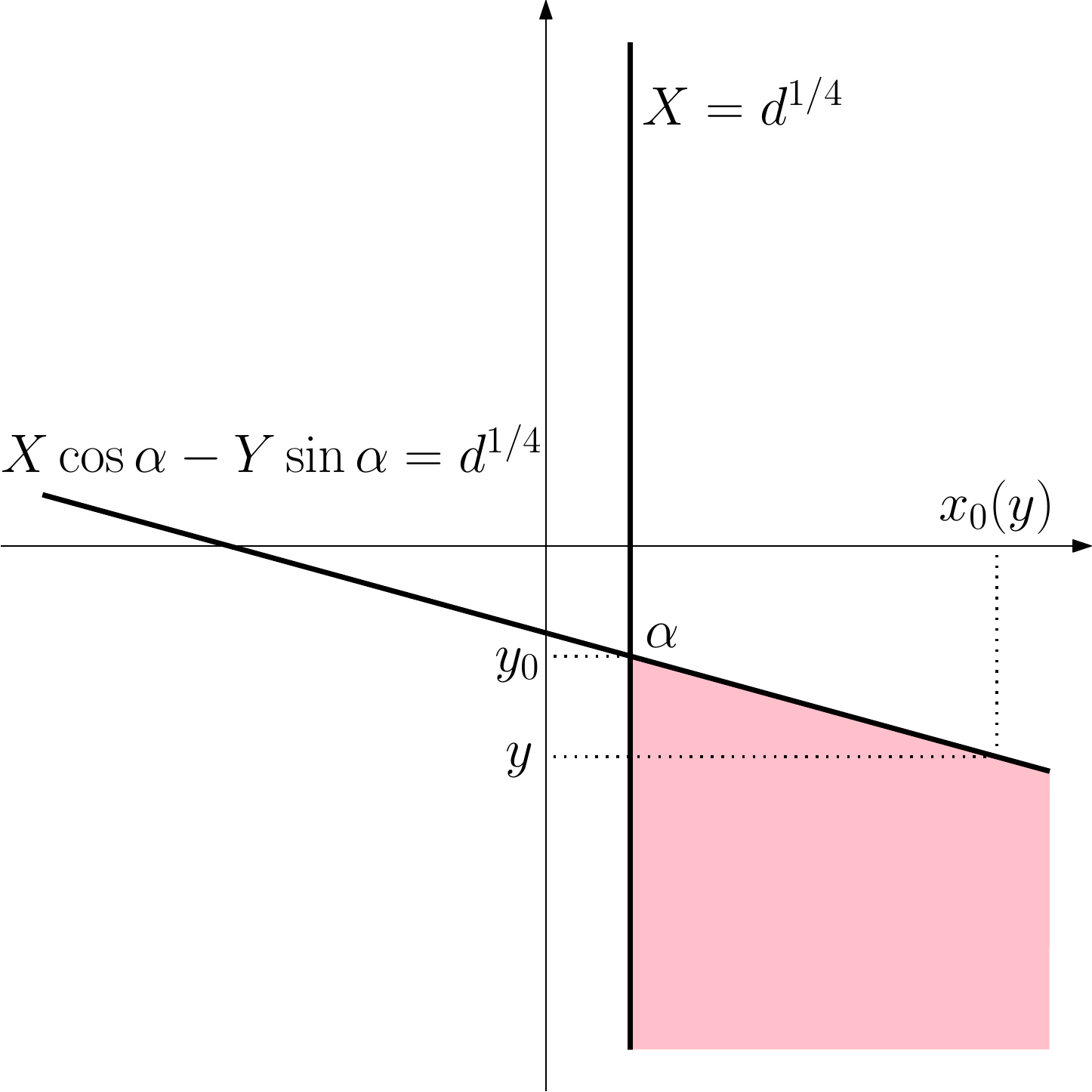}
        \caption{The case $\alpha > \pi / 2$}
    \end{subfigure}
    \end{center}
    \caption{The region corresponding to $X \geq d^{1/4}$
    and $X \cos \alpha - Y \sin \alpha \geq d^{1/4}$}
    \label{fig_area}
\end{figure}

Suppose that $u, v \in S^{d-1}$ are two points on the unit sphere
with angle $\alpha$ between them. Our goal is to estimate
the probability of collision
$\Prb{\Rc}{\Rc(u) = \Rc(v)}$, where $\Rc$ is a partition sampled
according to Spherical LSH. To compute the probability of collision,
observe that the probability that at a given iteration we ``capture''
\emph{either} $u$ or $v$ is equal to
$    \Prb{g \sim N(0, 1)^d}
    {\langle u, g \rangle \geq d^{1/4}
     \mbox{ or }\langle v, g \rangle \geq d^{1/4}}
    $. Similarly, the probability that we capture \emph{both} $u$ and
    $v$ is equal to
    $    \Prb{g \sim N(0, 1)^d}
    {\langle u, g \rangle \geq d^{1/4}
     \mbox{ and }\langle v, g \rangle \geq d^{1/4}}
    $. After a moment of thought, it becomes clear that the probability
    of collision $\Prb{\Rc}{\Rc(u) = \Rc(v)}$
    equals to the probability of the event
    ``both $u$ and $v$ are captured at a given iteration''
    conditioned on the event
    ``either $u$ or $v$ are captured at a given iteration''. Thus,
\begin{align}
    \Prb{\Rc}{\Rc(u) = \Rc(v)}
    & = \frac{
    \Prb{g \sim N(0, 1)^d}
    {\langle u, g \rangle \geq d^{1/4} 
     \mbox{ and }\langle v, g \rangle \geq d^{1/4}}
    }{
    \Prb{g \sim N(0, 1)^d}
    {\langle u, g \rangle \geq d^{1/4}
     \mbox{ or }\langle v, g \rangle \geq d^{1/4}}
    }
    \nonumber
    \\ & = \frac{
        \Prb{X, Y \sim N(0, 1)}{X \geq d^{1/4} 
        \mbox{ and } X \cos \alpha - Y \sin\alpha \geq d^{1/4}}
    }{
      \Prb{X, Y \sim N(0, 1)}{X \geq d^{1/4}\mbox{ or } X \cos \alpha - Y
      \sin \alpha \geq d^{1/4}}
    }
    \nonumber
    \\ & \in [0.5; 1] \cdot \frac{
        \Prb{X, Y \sim N(0, 1)}{X \geq d^{1/4} 
        \mbox{ and } X \cos \alpha - Y \sin\alpha \geq d^{1/4}}
    }{
        \Prb{X \sim N(0, 1)}{X \geq d^{1/4}}
    },\label{initial_expansion}
\end{align}
where the second step follows from the spherical
symmetry of Gaussians,
and the third step follows from the immediate equality
$$
\Prb{X \sim N(0, 1)}{X \geq d^{1/4}} = \Prb{X, Y \sim N(0, 1)}
 {X \cos \alpha - Y \sin\alpha \geq d^{1/4}}.
$$

By Lemma~\ref{gaussians},
\begin{equation}
\label{gaussian_estimate}
\Prb{X \sim N(0, 1)}{X \geq d^{1/4}} = (1 + O(d^{-1/2})) \cdot
\frac{e^{-\sqrt{d} / 2}}{(2 \pi)^{1/2} d^{1/4}},
\end{equation}
so, plugging~(\ref{gaussian_estimate}) into~(\ref{initial_expansion}),
we have
\begin{equation}
    \label{reduction}
    \Prb{\Rc}{\Rc(u) = \Rc(v)}
    = \Theta(d^{1/4}) \cdot \frac{
        \Prb{X, Y \sim N(0, 1)}{X \geq d^{1/4}
        \mbox{ and } X \cos \alpha - Y \sin\alpha \geq d^{1/4}}
    }{
        e^{-\sqrt{d} / 2}
    }.
\end{equation}
For $0 \leq \alpha \leq \pi$ and $\lambda > 0$ denote
$$
W_{\alpha, \lambda} := \set{(x, y) \colon x \geq \lambda \mbox{ and } x \cos \alpha - y \sin \alpha \geq \lambda} \subseteq \Rbb^2.
$$
Thus, estimating $\Prb{\Rc}{\Rc(u) = \Rc(v)}$
boils down to computing the Gaussian measure of $W_{\alpha, d^{1/4}}$ (see Figure~\ref{fig_area}).

For $d \in \Nbb$, $0 \leq \alpha \leq \pi$ denote
$$
    S(d, \alpha) := \Prb{X, Y \sim N(0, 1)}{(X, Y) \in W_{\alpha, d^{1/4}}}.
$$

\begin{claim}
    \label{monotone_bound}
    For every $d$ the function $S(d, \alpha)$ is non-increasing
    in $\alpha$.
\end{claim}
\begin{proof}
    One can check that for every $d \in \Nbb$ and
    $0 \leq \alpha' \leq \alpha'' \leq \pi$
    we have
    \begin{equation*}
        W_{\alpha', d^{1/4}} \supseteq W_{\alpha'', d^{1/4}},
    \end{equation*}
    hence the claim.
\end{proof}

Our next goal will be estimating $S(d, \alpha)$ for
$\alpha \in [d^{-\Omega(1)}; \pi - d^{-\Omega(1)}]$ within a factor
polynomial in $d$. \hl{Add heuristic explanation} 
We would like to claim that $S(d, \alpha)$ is close to
$\widetilde{S}(d, \alpha) := \Prb{X, Y \sim N(0, 1)}{X \geq d^{1/4} \mbox{ and }
Y \leq y_0}$, where $y_0 = -d^{1/4} \tan \frac{\alpha}{2}$
is the $y$-coordinate of the intersection of two lines:
$x = d^{1/4}$ and $x \cos \alpha - y \sin \alpha = d^{1/4}$
(see Figure~\ref{fig_area}).
But first let us compute $\widetilde{S}(d, \alpha)$---this
is easy due to the independence of $X$ and $Y$.

\begin{claim}
    \label{orth_estimate}
    If $\alpha = \Omega(d^{-1/5})$, then
    $$
        \widetilde{S}(d, \alpha)
        \in \frac{1 \pm d^{-\Omega(1)}}
        {2 \pi d^{1/2} \tan \frac{\alpha}{2}}
        \exp\left(-\left(1 + \tan^2 \frac{\alpha}{2}\right)\frac{\sqrt{d}}{2}\right).
    $$
\end{claim}
\begin{proof}
    First, observe that since $\alpha = \Omega(d^{-1/5})$,
    we have $y_0 = -d^{1/4} \tan \frac{\alpha}{2} = - d^{\Omega(1)}$.
    Next, we have
    \begin{align*}
        \widetilde{S}(d, \alpha) &= 
        \Prb{X,Y\sim N(0, 1)}{X \geq d^{1/4} \mbox{ and 
        } Y \leq y_0}
        \\&=
        \Prb{X \sim N(0, 1)}{X \geq d^{1/4}}
        \Prb{Y \sim N(0, 1)}{Y \leq y_0}
        \\&\in
        (1 \pm d^{-\Omega(1)}) \cdot
        \frac{e^{-\sqrt{d} / 2}}{\sqrt{2 \pi} d^{1/4}}
        \cdot
        \frac{e^{-|y_0|^2 / 2}}{\sqrt{2 \pi} |y_0|}
        \\&\in \frac{1 \pm d^{-\Omega(1)}}
        {2 \pi d^{1/2} \tan \frac{\alpha}{2}}
        \exp\left(-\left(1 + \tan^2 \frac{\alpha}{2}\right)\frac{\sqrt{d}}{2}\right),
    \end{align*}
    where the second step is by independence of $X$ and $Y$,
    the third step is due to $y_0 = -d^{\Omega(1)}$
    and Lemma~\ref{gaussians} and the fourth step is due to
    $y_0 = -d^{1/4} \tan \frac{\alpha}{2}$.
\end{proof}

Now the goal is to prove that if $\alpha$ is not too close
to $0$ or $\pi$, then $S(d, \alpha)$ is close to
$\widetilde{S}(d, \alpha)$.
\begin{claim}
    \label{error_bound}
    If $\alpha = \Omega(d^{-1/5})$ and
    $\alpha = \pi - \Omega(d^{-\delta})$ for a sufficiently small
    $\delta > 0$, then
    $$
        d^{-O(1)} \leq \frac{S(d, \alpha)}{\widetilde{S}(d, \alpha)} \leq d^{O(1)}.
    $$
\end{claim}
\begin{proof}
    First, for $\alpha = \pi / 2$ the claim is trivial, since
    $S(d, \pi / 2) = \widetilde{S}(d, \pi / 2)$.
    Next, we consider the cases $\alpha < \pi / 2$
    and $\alpha > \pi / 2$ separately (see Figure~\ref{fig_area}).

    \paragraph{The case $\alpha < \pi / 2$:}
    in this case we have
    $$
        S(d, \alpha) - \widetilde{S}(d, \alpha)
        = \frac{1}{2 \pi} \int_{y_0}^{\infty} \int_{x_0(y)}^{\infty}
        e^{-\frac{x^2 + y^2}{2}} \; dx \; dy,
    $$
    where $x_0(y) = \frac{d^{1/4} + y \sin \alpha}{\cos \alpha}$ (see Figure~\ref{fig_area}).
    For every $y \geq y_0$ we have $x_0(y) \geq d^{1/4}$, so
    by Lemma~\ref{gaussians}
    $$
        \int_{x_0(y)}^{\infty} e^{-x^2 / 2} dx
        \in \frac{(1 \pm d^{-\Omega(1)})e^{-x_0(y)^2 / 2}}{x_0(y)}.
    $$
    Thus,
    $$
        S(d, \alpha) - \widetilde{S}(d, \alpha) =
        \frac{1 \pm d^{-\Omega(1)}}{2 \pi} \int_{y_0}^{\infty}
        \frac{e^{-\frac{y^2 + x_0(y)^2}{2}}}{x_0(y)} \; dy.
    $$
    By direct computation,
    $$
        y^2 + x_0(y)^2 =
        y^2 + \left(\frac{d^{1/4} + y \sin \alpha}{\cos \alpha}\right)^2 = 
        \left(\frac{d^{1/4} \sin \alpha + y}{\cos \alpha}\right)^2 + d^{1/2}.
    $$
    Let us perform the following change of variables:
    $$
        u = \frac{d^{1/4} \sin \alpha + y}{\cos \alpha}.
    $$
    We have
    \begin{align*}
        S(d, \alpha) - \widetilde{S}(d, \alpha) & = 
        \frac{(1 \pm d^{-\Omega(1)})e^{-\sqrt{d} / 2}}{2 \pi}
        \int_{d^{1/4} \tan \frac{\alpha}{2}}^{\infty} \frac{e^{-u^2 / 2}}{d^{1/4} + u \tan \alpha} \; du
        \\ & \leq
        \frac{(1 \pm d^{-\Omega(1)})e^{-\sqrt{d} / 2}}{2 \pi d^{1/4}}
        \int_{d^{1/4} \tan \frac{\alpha}{2}}^{\infty} e^{-u^2 / 2} \; du
        \\ & =
        (1 \pm d^{-\Omega(1)}) \widetilde{S}(d, \alpha),
    \end{align*}
    where the last step is due to Lemma~\ref{gaussians} and Claim~\ref{orth_estimate}.
    Overall, we have
    $$
        \widetilde{S}(d, \alpha) \leq S(d, \alpha) \leq
        (2 \pm d^{-\Omega(1)}) \widetilde{S}(d, \alpha).
    $$

    \paragraph{The case $\alpha > \pi / 2$:} this case is similar.
    We have
    \begin{align*}
        \widetilde{S}(d, \alpha) - S(d, \alpha) &= 
        \frac{1}{2 \pi} \int_{-\infty}^{y_0} \int_{x_0(y)}^{\infty}
        e^{-\frac{x^2 + y^2}{2}} \; dx \; dy
        \\&=
        \frac{1 \pm d^{-\Omega(1)}}{2 \pi}
        \int_{-\infty}^{y_0} \frac{e^{-\frac{y^2 + x_0(y)^2}{2}}}{x_0(y)} \; dy.
    \end{align*}
    After the change
    $$
        u = \frac{d^{1/4} \sin \alpha + y}{\cos \alpha}
    $$
    we get (note that $\cos \alpha < 0$)
    \begin{align*}
        \widetilde{S}(d, \alpha) - S(d, \alpha) &=
        \frac{(1 \pm d^{-\Omega(1)}) e^{-\sqrt{d} / 2}}{2 \pi}
        \int_{d^{1/4} \tan \frac{\alpha}{2}}^{\infty} \frac{e^{-u^2 / 2}}{u |\tan \alpha| - d^{1/4}} \; du
        \\&\leq
        \frac{(1 \pm d^{-\Omega(1)}) e^{-\sqrt{d} / 2}}{2 \pi d^{1/4}(|\tan \alpha| \tan \frac{\alpha}{2} - 1)}
        \int_{d^{1/4} \tan \frac{\alpha}{2}}^{\infty} e^{-u^2 / 2} \; du
        \\&= \frac{1 \pm d^{-\Omega(1)}}{|\tan \alpha| \tan \frac{\alpha}{2} - 1} \widetilde{S}(d, \alpha).
    \end{align*}
    Since $\alpha = \pi - \Omega(d^{-\delta})$,
    we have
    $$
        \frac{1}{|\tan \alpha| \tan \frac{\alpha}{2} - 1}
        = \frac{1}{1 + \Omega(d^{-\delta})} = 1 - \Omega(d^{-\delta}).
    $$
    We can choose $\delta$ such that
    $$
        \widetilde{S}(d, \alpha) - S(d, \alpha)
        \leq (1 - d^{-O(1)}) \widetilde{S}(d, \alpha),
    $$
    so
    $$
        d^{-O(1)} \cdot \widetilde{S}(d, \alpha) \leq S(d, \alpha) \leq \widetilde{S}(d, \alpha).
    $$
\end{proof}

Combining (\ref{reduction}),
Claim~\ref{monotone_bound}, Claim~\ref{orth_estimate},
Claim~\ref{error_bound} and the formula
$$
    \tan^2 \frac{\alpha}{2} = \frac{\|u - v\|^2}{4 - \|u - v\|^2}, 
$$
we get the first two bullet points from the statement of
Theorem~\ref{gaussian_lsh_ideal}.

\subsection{Three-way collision probabilities}

In this section we prove~(\ref{glsh_cond}). We start with a simpler case $\eps = 0$.

\subsubsection{The case $\eps = 0$}
\label{eps_nil}

Suppose we have $u, v, w \in S^{d-1}$ with $\|u - v\| = \tau$ with $\tau \leq 1.99$,
$\|u - w\|, \|v - w\| = \sqrt{2}$.
Our goal is to upper bound
$$
    \Prb{\Rc}{\Rc(u) = \Rc(w) \mid \Rc(u) = \Rc(v)}.
$$
Similarly to the two-point case, we have
\begin{align*}
  & \Prb{\Rc}{\Rc(u) = \Rc(v) = \Rc(w)}
  \\ & = \frac{\Prb{g \sim N(0, 1)^d}{\langle u, g \rangle \geq d^{1/4}  \mbox{ and } \langle v, g \rangle \geq d^{1/4} \mbox{ and } \langle w, g \rangle \geq d^{1/4}}}{\Prb{g \sim N(0, 1)^d}{\langle u, g \rangle \geq d^{1/4}  \mbox{ or } \langle v, g \rangle \geq d^{1/4} \mbox{ or } \langle w, g \rangle \geq d^{1/4}}}
 \\ & =
    \frac{\Prb{X,Y,Z \sim N(0, 1)}
    {X \geq d^{1/4} \mbox{ and } Y \sqrt{1 - \frac{\tau^2}{4}} + Z \cdot \frac{\tau}{2} \geq d^{1/4}
    \mbox{ and } Y \sqrt{1 - \frac{\tau^2}{4}} - Z \cdot \frac{\tau}{2} \geq d^{1/4}}}{\Prb{X,Y,Z \sim N(0, 1)}
    {X \geq d^{1/4} \mbox{ or } Y \sqrt{1 - \frac{\tau^2}{4}} + Z \cdot \frac{\tau}{2} \geq d^{1/4}
    \mbox{ or } Y \sqrt{1 - \frac{\tau^2}{4}} - Z \cdot \frac{\tau}{2} \geq d^{1/4}}}
 \\ & = \Theta(1) \cdot
    \frac{\Prb{X,Y,Z \sim N(0, 1)}
    {X \geq d^{1/4} \mbox{ and } Y \sqrt{1 - \frac{\tau^2}{4}} + Z \cdot \frac{\tau}{2} \geq d^{1/4}
      \mbox{ and } Y \sqrt{1 - \frac{\tau^2}{4}} - Z \cdot \frac{\tau}{2} \geq d^{1/4}}}{\Prb{X \sim N(0, 1)}{X \geq d^{1/4}}}
 \\ & = \Theta(1) \cdot
    \Prb{Y,Z \sim N(0, 1)}
    {Y \sqrt{1 - \frac{\tau^2}{4}} + Z \cdot \frac{\tau}{2} \geq d^{1/4}
      \mbox{ and } Y \sqrt{1 - \frac{\tau^2}{4}} - Z \cdot \frac{\tau}{2} \geq d^{1/4}}
    \\ & = \Theta(1) \cdot \Prb{X \sim N(0, 1)}{X \geq d^{1/4}} \cdot \Prb{\Rc}{\Rc(u) = \Rc(v)},
\end{align*}
where the first step is similar to the two-point case,
the second step is by the spherical symmetry of Gaussians,
the third step is due to the following immediate identity
\begin{multline*}
\Prb{X \sim N(0, 1)}{X \geq d^{1/4}}
=
\Prb{Y, Z \sim N(0, 1)}{Y \sqrt{1 - \frac{\tau^2}{4}} + Z \cdot \frac{\tau}{2} \geq d^{1/4}} \\ = \Prb{Y, Z \sim N(0, 1)}{Y \sqrt{1 - \frac{\tau^2}{4}} - Z \cdot \frac{\tau}{2} \geq d^{1/4}},
\end{multline*}
  the fourth step is due to the independence of $X$, $Y$ and $Z$,
  and the last step is due to~(\ref{initial_expansion}).
Thus,
$$
    \Prb{\Rc}{\Rc(u) = \Rc(w) \mid \Rc(u) = \Rc(v)} = \Theta(1) \cdot \Prb{X \sim N(0, 1)}{X \geq d^{1/4}},
$$
which, combined with Lemma~\ref{gaussians}, gives the desired claim.

\subsubsection{The case of arbitrary $\eps$}

Suppose that $u, v, w \in S^{d-1}$ are such that $\|u - v\| = \tau$ with $\tau \leq 1.99$,
$\|u - w\|, \|v - w\| \in \sqrt{2} \pm \eps$ with $\eps = o(1)$.
We would like to show~(\ref{glsh_cond}) for this case. In a nutshell, the goal is
to show that the bound proved in Section~\ref{eps_nil} is stable under small perturbations in $\eps$.

We are interested in lower bounding
\begin{equation}
  \label{random_1}
\ln \frac{1}{\Prb{\Rc}{\Rc(u) = \Rc(w) \mid \Rc(u) = \Rc(v)}}
= \ln \frac{1}{\Prb{\Rc}{\Rc(u) = \Rc(v) = \Rc(w)}} - \ln \frac{1}{\Prb{\Rc}{\Rc(u) = \Rc(v)}}.
\end{equation}

First, if $\tau \leq \eps^{\nu}$, where $\nu > 0$ is a small positive constant (to be chosen later),
then we can proceed as in~(\ref{naive_cond}): we use that
\begin{equation}
  \label{random_2}
\ln \frac{1}{\Prb{\Rc}{\Rc(u) = \Rc(v) = \Rc(w)}} \geq \ln \frac{1}{\Prb{\Rc}{\Rc(u) = \Rc(w)}}
\geq (1 - \eps^{\Omega(1)} - d^{-\Omega(1)}) \cdot \frac{\sqrt{d}}{2},
\end{equation}
due to $\|u - w\| \geq \sqrt{2} - \eps$ and~(\ref{glsh_lb}). On the other hand,
\begin{equation}
  \label{random_3}
\ln \frac{1}{\Prb{\Rc}{\Rc(u) = \Rc(v)}} \leq (\eps^{\Omega_{\nu}(1)} + d^{-\Omega(1)}) \cdot \sqrt{d},
\end{equation}
due to $\|u - v\| \leq \eps^{\nu}$ and~(\ref{glsh_ub}). Thus, combining~(\ref{random_1}), (\ref{random_2}) and~(\ref{random_3}), we are done.

Thus, we can assume that $\tau \geq \eps^{\nu}$ for a small $\nu > 0$. Due to the spherical symmetry of Gaussians,
we can assume that $u, v, w \in S^2$. Let $u', v', w' \in S^2$ be such that $\|u' - w'\| = \|v' - w'\| = \sqrt{2}$ and $\|u' - v'\| = \tau$.
From Section~\ref{eps_nil} we know that
$$
\ln \frac{1}{\Prb{\Rc}{\Rc(u') = \Rc(w') \mid \Rc(u') = \Rc(v')}} \geq (1 - d^{-\Omega(1)}) \cdot \frac{\sqrt{d}}{2},
$$
besides that,
$$
\ln \frac{1}{\Prb{\Rc}{\Rc(u) = \Rc(v)}} = \ln \frac{1}{\Prb{\Rc}{\Rc(u') = \Rc(v')}},
$$
since $\|u - v\| = \|u' - v'\| = \tau$; thus, it is sufficient to show that
\begin{equation}
  \label{yalocalgoal}
\ln \frac{1}{\Prb{\Rc}{\Rc(u) = \Rc(v) = \Rc(w)}}  \geq
\ln \frac{1}{\Prb{\Rc}{\Rc(u') = \Rc(v') = \Rc(w')}} - (\eps^{\Omega_\nu(1)} + d^{-\Omega(1)}) \cdot \sqrt{d},
\end{equation}
provided that $\nu > 0$ is small enough.
Recall that
\begin{align}
  \label{comp_1}
  \Prb{\Rc}{\Rc(u) = \Rc(v) = \Rc(w)} & =
  \frac{\Prb{g \sim N(0, 1)^3}{\langle u, g\rangle \geq d^{1/4} \wedge
      \langle v, g\rangle \geq d^{1/4}\wedge \langle w, g\rangle \geq d^{1/4}}}
       {\Prb{g \sim N(0, 1)^3}{\langle u, g\rangle \geq d^{1/4} \vee
           \langle v, g\rangle \geq d^{1/4}\vee \langle w, g\rangle \geq d^{1/4}}}\\
       \label{comp_2}
       \Prb{\Rc}{\Rc(u') = \Rc(v') = \Rc(w')} & =
       \frac{\Prb{g \sim N(0, 1)^3}{\langle u', g\rangle \geq d^{1/4} \wedge
           \langle v', g\rangle \geq d^{1/4}\wedge \langle w', g\rangle \geq d^{1/4}}}
            {\Prb{g \sim N(0, 1)^3}{\langle u', g\rangle \geq d^{1/4} \vee
                \langle v', g\rangle \geq d^{1/4}\vee \langle w', g\rangle \geq d^{1/4}}}.
\end{align}
Observe that the denominators in~(\ref{comp_1}) and~(\ref{comp_2}) are within a factor $3$ from each other.
Thus, it is sufficient to prove that
\begin{multline*}
  \ln \frac{1}{\Prb{g \sim N(0, 1)^3}{\langle u, g\rangle \geq d^{1/4} \wedge
      \langle v, g\rangle \geq d^{1/4}\wedge \langle w, g\rangle \geq d^{1/4}}}
    \\ \geq
  \ln \frac{1}{\Prb{g \sim N(0, 1)^3}{\langle u', g\rangle \geq d^{1/4} \wedge
      \langle v', g\rangle \geq d^{1/4}\wedge \langle w', g\rangle \geq d^{1/4}}}
  - (\eps^{\Omega_{\nu}(1)} + d^{-\Omega(1)}) \cdot \sqrt{d},
\end{multline*}
provided that $\nu > 0$ is small enough.

The joint distribution of $\langle u', g\rangle$, $\langle v', g \rangle$ and $\langle w', g \rangle$
is a multivariate Gaussian with zero mean and covariance matrix
$$
\mathcal{D} = \left(\begin{array}{ccc}
  1 & 1 - \frac{\tau^2}{2} & 0 \\
  1 - \frac{\tau^2}{2} & 1 & 0 \\
  0 & 0 & 1
\end{array}\right).
$$
Similarly, for $\langle u, g\rangle$, $\langle v, g \rangle$ and $\langle w, g \rangle$
the mean is zero and the covariance matrix is
$$
\mathcal{C} = \left(\begin{array}{ccc}
  1 & 1 - \frac{\tau^2}{2} & \pm O(\eps) \\
  1 - \frac{\tau^2}{2} & 1 & \pm O(\eps) \\
  \pm O(\eps) & \pm O(\eps) & 1
\end{array}\right).
$$

Observe that both $\mathcal{C}$ and $\mathcal{D}$ have all eigenvalues being at least
$\eps^{O(\nu)}$ and at most $O(1)$ by Gershgorin's theorem and due to the bound $\tau \geq \eps^{\nu}$.
In particular, both $\mathcal{C}$ and $\mathcal{D}$ are invertible.

\begin{definition}
  \label{mu_notation}
  For a closed subset $U \subseteq \Rbb^d$ we denote $\mu(U)$ the (Euclidean) distance from $0$ to $U$.
\end{definition}

In order to prove~(\ref{yalocalgoal}) we need to show that the probability that the centered Gaussian
with covariance matrix $\mathcal{C}$ belongs to the set
$T = \set{(x, y, z) \in \Rbb^3 : x \geq d^{1/4}, y \geq d^{1/4}, z \geq d^{1/4}}$ is not much larger
than the same probability for the centered Gaussian with covariance matrix $\mathcal{C}'$.
Using the results of Section~\ref{perturbed_section}, we get
\begin{multline}
  \label{blah_1}
  \Prb{g \sim N(0, 1)^3}{\langle u, g\rangle \geq d^{1/4} \wedge
    \langle v, g\rangle \geq d^{1/4}\wedge \langle w, g\rangle \geq d^{1/4}}
  \\ =
\Prb{x \sim N(0, \mathcal{C})}{x \in T}
\leq
O(1) \cdot \frac{e^{-\mu(\mathcal{C}^{-1/2} T)^2/2}}{\mu(\mathcal{C}^{-1/2} T)}
\leq O(1) \cdot
\frac{
  (1 + \eps^{\Omega_{\nu}(1)})
  e^{-(1 - \eps^{\Omega_{\nu}(1)}) \mu(\mathcal{D}^{-1/2} T)^2}}{\mu(\mathcal{D}^{-1/2} T)}.
\end{multline}
Now observe that by the results of Section~\ref{eps_nil} we have
\begin{equation}
  \label{blah_2}
  \Prb{g \sim N(0, 1)^3}{\langle u', g\rangle \geq d^{1/4} \wedge
    \langle v', g\rangle \geq d^{1/4}\wedge \langle w', g\rangle \geq d^{1/4}}
  = e^{-(1 \pm d^{\Omega(1)}) \mu(\mathcal{D}^{-1/2} T)^2 / 2}.
  \end{equation}
  Combining~(\ref{blah_1}) and~(\ref{blah_2}), we are done.

\subsection{Efficiency}
\label{sec_eff}

As has been already observed, the partitioning scheme as described in Section~\ref{sec_glsh}
is not efficient. One can fix this issue as follows.
Instead of checking, whether $\bigcup \Rc \ne S^{d-1}$, we can just stop after $T$ iterations.
If we choose $T$ such that the probability of the event ``$\bigcup \Rc \ne S^{d-1}$'' after $T$ steps
is less than $e^{-d^{100}}$, then we are done, since the bounds stated in Theorem~\ref{gaussian_lsh_ideal}
would remain true.

For a fixed point $u \in S^{d-1}$ we have
$$
    \Prb{g \sim N(0, 1)^d}{\langle u, g \rangle \geq d^{1/4}}
    = \Prb{X \sim N(0, 1)}{X \geq d^{1/4}} \geq e^{-O(\sqrt{d})},
$$
where the first step is by $2$-stability of Gaussians and the second step is by Lemma~\ref{gaussians}.

Now it is not hard to see taking the union bound over a sufficiently fine $\eps$-net that
if we set $T = e^{O(\sqrt{d})}$, then we get what we want.
This way, we conclude the time and space bounds in Theorem~\ref{gaussian_lsh_ideal}.

\section{Perturbed Gaussian measures}

\label{perturbed_section}

Let $A \in \Rbb^{d \times d}$ be a symmetric positive-definite matrix. Consider a centered $d$-variate Gaussian $N(0, A)$,
whose covariance matrix equals to $A$, and let $S \subseteq \Rbb^d$ be a closed \emph{convex} set.
\begin{multline}
  \label{estimate_ga}
\Prb{x \sim N(0, A)}{x \in S} = \Prb{y \sim N(0, I)}{A^{1/2} y \in S} =
\Prb{y \sim N(0, I)}{y \in A^{-1/2} S} \\ \leq \Prb{z \sim N(0, 1)}{z \geq d(0, A^{-1/2} S)}
\leq \frac{1}{\sqrt{2 \pi}} \cdot \frac{1}{\mu(A^{-1/2} S)} \cdot e^{-\mu(A^{-1/2} S)^2 / 2},
\end{multline}
where the first step is due to properties of multivariate Gaussians, the third step is due to the spherical
symmetry of $N(0, I)$ and due to the convexity of $S$ (and hence $A^{-1/2} S$); the last step is due to~(\ref{gaussians}).

Now suppose that $A \succeq \lambda I$, where $\lambda > 0$, and $B \in \Rbb^{d \times d}$ is such that $\|B - A\| \leq \eps$
for some $\eps > 0$ with $\eps \ll \lambda$.
Clearly, similarly to~(\ref{estimate_ga}), we get
\begin{equation}
  \label{estimate_gb}
\Prb{x \sim N(0, B)}{x \in S} \leq \frac{1}{\sqrt{2 \pi}} \cdot \frac{1}{\mu(B^{-1/2} S)} \cdot e^{-\mu(B^{-1/2} S)^2 / 2}.
\end{equation}

\begin{definition}
  Two matrices $A$ and $B$ are called $\eps$-spectrally close, if
  $$
  e^{-\eps} \cdot A \preceq B \preceq e^{\eps} \cdot A.
  $$
\end{definition}

We would like to claim that the right-hand sides of~(\ref{estimate_ga}) and~(\ref{estimate_gb}) are quite close.
For this it is sufficient to compare
$\mu(A^{-1/2} S)$ and $\mu(B^{-1/2} S)$.
Since
$$
\|A^{-1} B - I\| \leq \|A^{-1}\| \|B - A\| \leq \frac{\eps}{\lambda},
$$
We get that $A$ and $B$ are $O(\eps / \lambda)$-spectrally close. Thus, $A^{-1}$ and $B^{-1}$ are $O(\eps/\lambda)$-spectrally close.
Finally, $A^{-1/2}$ and $B^{-1/2}$ are $O(\eps / \lambda)$-spectrally close as well.
Thus,
$$
\mu(B^{-1/2} S) \in \left(1 \pm O\left(\frac{\eps}{\lambda}\right)\right) \cdot \mu(A^{-1/2} S),
$$
and
$$
\Prb{x \sim N(0, B)}{x \in S} \leq \frac{1}{\sqrt{2 \pi}} \cdot \frac{1 \pm O(\eps / \lambda)}{\mu(A^{-1/2} S)} \cdot
e^{-(1 \pm O(\eps / \lambda)) \cdot  \mu(A^{-1/2} S)^2 / 2}.
$$

\end{document}